\documentclass[journal,draftcls,onecolumn,12pt,twoside]{IEEEtranTCOM}
\usepackage{tikz}
\usepackage{acronym}
\usetikzlibrary{shapes,snakes}
\usepackage{empheq}
\usepackage{graphicx}
\usepackage{epstopdf}
\usepackage{multirow}
\usepackage{tabu}
\usepackage{array}
\usepackage{url}
\usepackage{amsmath}
\usepackage{cleveref}
\usepackage{amssymb}
\usepackage{amsthm}
\usepackage{epstopdf}
\usepackage{amsfonts}
\usepackage{tikz}
\usepackage{bm}
\usepackage{algorithmic}
\usepackage[ruled,vlined,linesnumbered]{algorithm2e}
\usetikzlibrary{arrows}
\usepackage{cite}
\usepackage{caption}
\usepackage{color}
\usepackage{subcaption}
\usepackage{amssymb}
\usepackage{tabulary}
\usepackage{booktabs}

\usepackage[justification=centering]{caption} 
\DeclareCaptionLabelFormat{algnonumber}{algorithm}
\newcommand{\rev}[1]{\textcolor{black}{#1}}

\tikzstyle{int}=[draw, fill=white!20, minimum size=2em]
\tikzstyle{init} = [pin edge={to-,thin,black}]

\newtheorem{lm}{Lemma}
\newtheorem{thm}{Theorem}

\newcommand\blfootnote[1]{%
	\begingroup
	\renewcommand\thefootnote{}\footnote{#1}%
	\addtocounter{footnote}{-1}%
	\endgroup
}

 \DeclareRobustCommand*{\IEEEauthorrefmark}[1]{%
 	\raisebox{0pt}[0pt][0pt]{\textsuperscript{\footnotesize\ensuremath{#1}}}}

\acrodef{AoI}{Age of Information}
\acrodef{IoT}{Internet of Things}
\acrodef{DPP}{Drift-Plus-Penalty}
\acrodef{DTMC}{Discrete-Time Markov Chain}
\acrodef{w.p.}{with probability}
\acrodef{EVD}{EigenValue-Decomposition}
\acrodef{CMDP}{Constrained Markov Decision Process}
\acrodef{FIFO}{First-in-First-Out}
\acrodef{MDP}{Markov Decision Process}
\acrodef{AoII}{Age of  Incorrect Information}
\acrodef{OFRP}{Old-or-Fresh Randomized Policy}
\acrodef{FoRP}{Fresh-only Randomized Policy}
\acrodef{w.p.}{with probability}

\IEEEoverridecommandlockouts                          

\title{
  Joint Sampling and Transmission  Policies for Minimizing Cost under AoI Constraints 
}

\author{
	\IEEEauthorblockN{Emmanouil Fountoulakis, 
		Marian Codreanu, \textit{Member, IEEE}, Anthony Ephremides, \textit{Life Fellow, IEEE},
	and Nikolaos Pappas, \textit{Senior Member, IEEE}}
}

\begin{document}
\immediate\write18{echo $PATH > tmp1}
\immediate\write18{/Library/TeX/texbin/epstopdf > tmp2} 

\maketitle
\thispagestyle{plain} 
\pagestyle{plain}

\begin{abstract}
\blfootnote{This paper extends the work in \cite{FountoulakisOptimalSampling}.
	
E. Fountoulakis is with the Communication Systems Department, EURECOM, Sophia Antipolis, France. Email: emmanouil.fountoulakis@eurecom.fr.
M. Codreanu is with the Department of Science and Technology, Linköping University, Norrköping, Sweden.  Email:  marian.codreanu@liu.se.
N. Pappas is with the Department of Computer and Information Science, Liköping University, Linköping, Sweden. Email: nikolaos.pappas@liu.se.
A. Ephremides is with the Electrical and Computer Engineering Department, University of Maryland, College Park, USA. Email: \{etony@umd.edu\}.}
In this work, we consider the problem of jointly minimizing the average cost of sampling and
transmitting status updates by users over a wireless channel subject to average Age of Information (AoI)
constraints. Errors in the transmission may occur and a policy has to decide if the users
sample a new packet or attempt to retransmission the packet sampled previously. The cost consists
of both sampling and transmission costs. The sampling of a new packet after a failure imposes an
additional cost on the system. We formulate a stochastic optimization problem with the average cost in
the objective under average AoI constraints. To solve this problem, we propose three scheduling policies;
a) a dynamic policy, that is centralized and requires full knowledge of the state of the system, b) two
stationary randomized policies that require no knowledge of the state of the system. We utilize tools
from Lyapunov optimization theory and Discrete-Time Markov Chain (DTMC) to provide the dynamic policy and the randomized ones, respectively. Simulation results show the importance of providing
the option to transmit an old packet in order to minimize the total average cost.
\end{abstract}

\begin{IEEEkeywords}
Age of Information (AoI), Lyapunov optimization, randomized policies, scheduling, wireless networks
\end{IEEEkeywords}
	
\IEEEpeerreviewmaketitle 
	
\section{Introduction}
The \ac{AoI} is a metric that captures the timeliness or freshness of the data \cite{kostamonograph2017age,sunmodiano2019age,yates2020age}. It was first introduced in \cite{kaulyates2012real}, and it is defined as the time elapsed since the generation of the status update that was most recently received by a destination.
AoI can play an important role in applications with freshness-sensitive data, e.g., environment
monitoring, smart agriculture, sensor networks, etc. Consider a cyber-physical system, where a number of sensors sample and transmit freshness-sensitive data (e.g., temperature, humidity, solar radiation level) to a destination over a wireless channel. However, devices who sample fresh information could perform more sophisticated tasks rather than simply sample new information for transmitting it to the destination. For example, an \ac{IoT} device can perform initial feature extraction and pre-classification by using machine learning tools. In such cases, the sampling cost cannot be ignored, and especially for low power budget wireless devices. In this work, we consider a set of users who sample and transmit fresh information over error-prone channels. Our goal is to provide scheduling policies that minimize the transmission and sampling cost of the communications system while satisfying the average \ac{AoI} requirements. We address the  trade-off between average \ac{AoI} and total cost. Simulation results show for which cases it is  beneficial to send an old yet not stale packet for minimizing the total cost while providing the required freshness of the data at the destination.
\subsection{Related Works}
The average and peak \ac{AoI} analysis in queueing systems has been extensively studied over the last few years, \cite{kamephr2015effect,kosta2019queuemanage, talak2020age}. The studies mainly focus on the derivation of closed-form or approximated expressions of the average and peak \ac{AoI} in different network set-ups and under different queueing disciplines.  Also, \ac{AoI} performance analysis has been studied  in random-access networks, \cite{ pappas2019delay, fountoulakis2020information}, as well as in a CSMA environment \cite{maatouk2020age}. In \cite{pappas2019delay,fountoulakis2020information}, the authors study the interplay between \ac{AoI}-oriented users and delay-constrained users in random-access networks.

Optimization and control of wireless networks with \ac{AoI}-oriented users has been recently investigated by the research community for a plethora of network scenarios,  \cite{kadota2021minimizing,joo2018wireless,qian2020minimizing, talak2020improving, chen2019optimal, moltafet2019power, tripathi2019whittle, ceran2019reinforcement, abd2020aoi,zhou2019joint,StamatakisIoT,kadota2019scheduling,BedewyOptimalSampling,yao2020age}. 
The authors, in \cite{kadota2021minimizing,joo2018wireless,qian2020minimizing}, study single-hop networks with stochastic arrivals, in which the status updates randomly  arrive to the users' queues and wait for being transmitted to the destination. Max-weight policies, as well as stationary randomized policies, are provided for minimizing the \ac{AoI}.
On the other hand, scenarios, in which \textit{generate-at-will} policy is utilized, have been considered in the literature, \cite{kadota2019scheduling,chen2019optimal,moltafet2019power,tripathi2019whittle,ceran2019reinforcement, zhou2019joint,StamatakisIoT, abd2020aoi}. In these scenarios, which are closer to our work, no random arrivals are considered, and the transmitters generate status updates from a source based on their or  scheduler's decisions.
 In \cite{kadota2019scheduling}, the problem of \ac{AoI} minimization with throughput constraints in a wireless network with multiple users is considered. The authors provide a lower bound for the average \ac{AoI}. Several scheduling policies for minimizing the \ac{AoI} are proposed. Furthermore, the \ac{AoI} minimization problem with heterogeneous traffic has been studied in \cite{chen2019optimal, fountoulakis2021dynamic}, and the power minimization under \ac{AoI} constraints in \cite{moltafet2019power}. 
General non-decreasing cost functions of \ac{AoI} has been considered for the \ac{AoI} minimization problem in \cite{tripathi2019whittle}. In \cite{BedewyOptimalSampling}, the authors consider the joint sampling and scheduling problem for minimizing \ac{AoI} in multi-source systems. In \cite{yao2020age}, the authors consider the \ac{AoI} minimization problem with average energy consumption constraints  of which the optimal policy is shown to be a randomized mixture 
two stationary deterministic policies. 

In \cite{ceran2019reinforcement,StamatakisIoT, zhou2019joint, abd2020aoi}, the optimization of \ac{AoI} in \ac{IoT} and energy harvesting communication systems has been studied.
In \cite{ceran2019reinforcement,abd2020aoi}, the authors formulate the \ac{AoI} minimization problem for scenarios, in which energy-harvesting nodes with finite battery capacity sample fresh information and transmit it  to the destination. The problems are formulated as \ac{MDP} and they are solved by using tools from dynamic programming and reinforcement learning.  In \cite{StamatakisIoT,zhou2019joint}, the authors consider the \ac{AoI} minimization in scenarios in which \ac{IoT} devices with heterogeneous traffic and limited energy budget, respectively, sample and transmit fresh information.

\ac{AoI} has been also considered an important metric for remote estimation \cite{huang2020real,kam2020age,inoue2019aoi,ornee2019sampling,huang2019retransmit}, and a first step towards semantics-aware communication systems, \cite{kountouris2020semantics,maatouk2020ageSemantics,popovski2020semantic}. 
In \cite{huang2020real}, the authors address the trade-off between reliability and freshness of information in a wireless sensor system. In \cite{kam2020age}, the metric of \ac{AoII} is introduced for remote estimation. In \cite{inoue2019aoi}, the authors consider continuous-time Markovian sources observed by a remote monitor.  In \cite{ornee2019sampling}, the authors design an optimal sampler for remote estimation of random processes by utilizing the framework of continuous \ac{MDP}. In \cite{huang2019retransmit}, the authors study the \ac{AoI} as a performance metric for remote estimation in a wireless networked control system.  

The scenario that is closer to our work is the one considered in \cite{zhou2019joint}. The authors consider an \ac{IoT} device sampling and sending information over a wireless fading channel to a receiver. Sampling, as well as the transmission cost, are considered, and the power transmission is adapted according to the channel conditions for ensuring reliable communication. The authors formulate the \ac{AoI} minimization problem with energy constraints as a \ac{CMDP}, and they propose a structural-aware optimal policy for solving the \ac{CMDP} problem. In our work,  the goal is different, namely the cost minimization under average \ac{AoI} constraints. Furthermore, we consider error-prone wireless channels, an assumption that makes our problem fundamentally different. In our case, the transmitters have also the option to retransmit an old packet, and it is shown that this option can dramatically improve  the system cost while guaranteeing the required data freshness at the receiver, for each user.

\subsection{Contributions} 
In this work, we consider the minimization of the total average cost while guaranteeing average \ac{AoI} below a threshold for each user. We propose three scheduling policies. The first scheduling policy, named \ac{DPP}, is a dynamic policy that takes decisions in each slot to minimize the cost. However, the \ac{DPP} policy requires full information, i.e., the cache state of each user, and the waiting time of the corresponding packet. For this reason, we propose two stationary randomized policies that require no information. The first stationary randomized policy, named \ac{OFRP}, allows the scheduler to decide probabilistically at every time slot the scheduled user. Then, the scheduled user decides by itself the action, i.e., either to sample and transmit, or to transmit an old packet, or to remain idle. The second stationary policy,  named \ac{FoRP}, that is a simplified version of the first, allows the scheduler to decide probabilistically which user to schedule at every time slot. Then the scheduled user decides either to sample and transmit or to remain idle. 

The contributions of our work are summarized below:
\begin{itemize}
	\item We propose a dynamic policy  based on Lyapunov optimization, and we prove that its solution is near-optimal. 
	\item We propose two stationary randomized policies. We model the system as two \acp{DTMC}, and we provide the expressions for the total average cost, the average \ac{AoI} for each user, as well as, the distribution of the \ac{AoI}. 
	\item Simulations results are provided, and they  show the performance of the proposed scheduling policies.
	Also, simulation results  show how the option of transmitting an old packet can dramatically improve  the total average cost. 
\end{itemize}

\section{System Model}\label{Sec:SytemModel}

\begin{figure}[t!]
	\centering
	\includegraphics[scale=0.17]{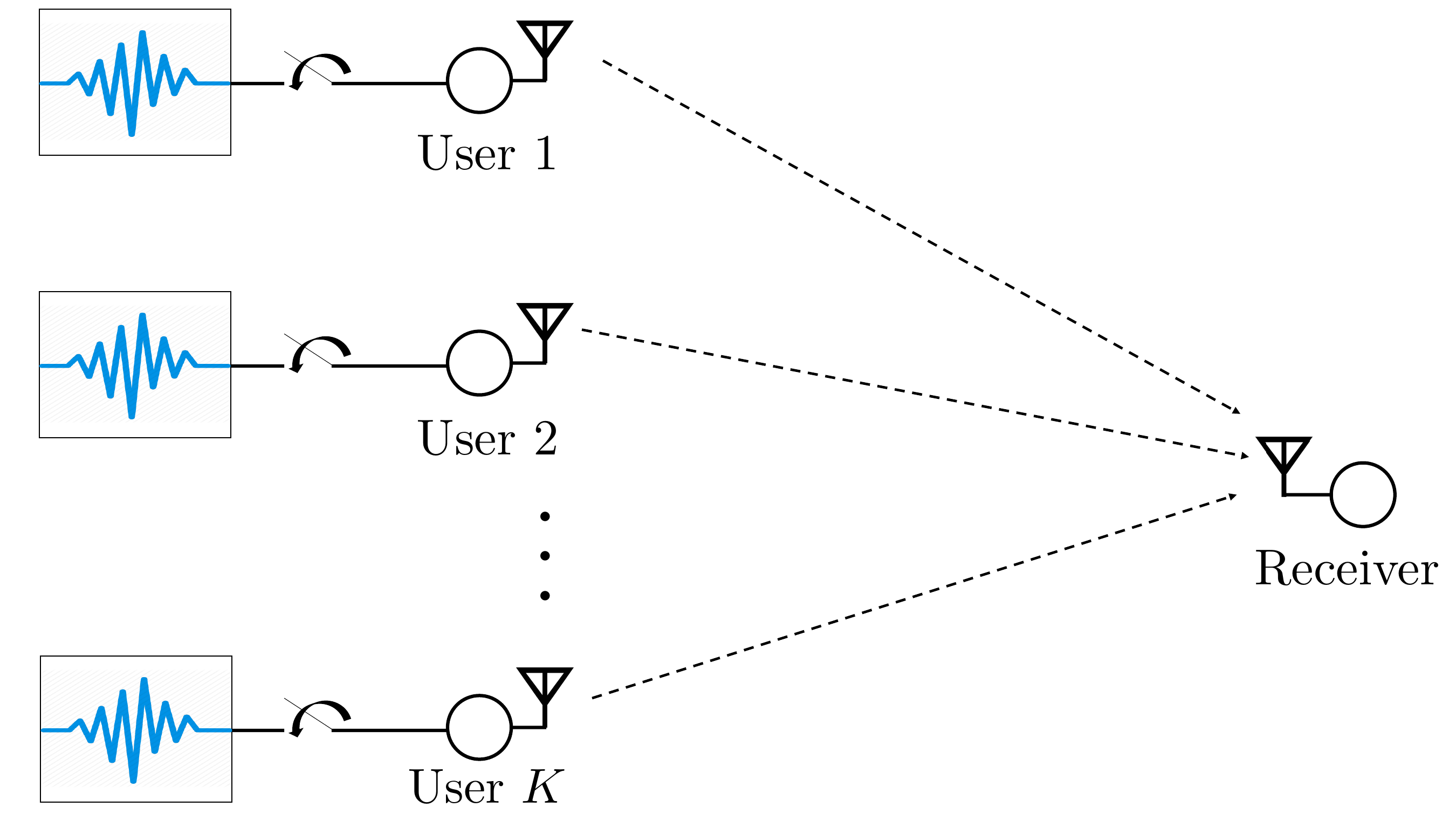}
	\caption{System model.}
	\label{fig:systemmodel}
\end{figure}

We consider a set of users, denoted by $\mathcal{K} = \left\{1,\ldots, K\right\}$, who sample fresh information and send this information, in form of packets, to a receiver over a wireless fading channel, as shown in Fig. \ref{fig:systemmodel}. Time is assumed to be slotted. Let $t \in \mathbb{Z}_+$ be the $t^{\text{th}}$ slot. Note that due to fluctuations of the fading channel we may have error in the transmissions. Therefore, the packet is transmitted successfully to the receiver with some  probability. Note that in the case of error transmission, the user keeps the packet in its cache for possible retransmission during the next slots. 

We consider that at every time slot, up to one user is scheduled to  transmit a packet.
Let $p_{k}$ be the success transmission probability of user $k$.
We denote by $Q_{k}(t)$ the state of the cache of user $k$. $Q_{k}(t)$ takes the value of $1$, if there is a packet in the cache, and $0$ otherwise.
We denote by $s_k(t)$ the action of the user $k$  to sample and transmit in time slot $t$, where
\begin{align}
		s_{k}(t) = 
		\begin{cases}
			 	1\text{, if the user } k \text{ samples and transmits in time slot }t\text{,}\\
				0\text{, otherwise.}
		\end{cases}
		\label{eq: sampleindicator}
\end{align}
 We denote by $\mu_{k}(t)$ the action of  user $k$ to transmit an old packet in time slot $t$, where
 \begin{align}
 	\mu_k(t) = 
 	\begin{cases}
 		1\text{, if the user } k \text{ transmits an old packet in time slot } t\text{,}\\
 		0\text{, otherwise.}
 	\end{cases}
 	\label{eq: oldpacketindicator}
 \end{align}
 Note that $\mu_k(t)$ can take the value of one only if there is packet in the cache. 
 We also denote by $d_{k}(t)$ the successful packet transmission of user $k$, where
 \begin{align}
 		d_{k}(t) = 
 		\begin{cases}
 					1\text{, if the receiver receives a packet from user }  k\\ \text{\quad during the time slot } t-1\text{,}\\
 					0\text{, otherwise.}  
		\end{cases}
		\label{eq: packetrecindicator}
 \end{align}
Note that $d_{k}(t)$ takes the value of one, if  $s_k(t-1)=1$ or $\mu_k(t-1)=1$, with probability $p_k$, and 
$s_k(t) + \mu_k(t) \leq 1\text{, } \forall k$.
It follows that $\mathbb{E} \{d_k(t) | \mu_k(t-1), s_k(t-1)\} = p_k\mu_k(t-1)  + p_k s_k(t-1)$. By applying the law of iterated expectations, we obtain$
	\mathbb{E} \{d_k (t) \} = p_k\mathbb{E} \{\mu_k(t-1)\}  + p_k \mathbb{E} \{ s_k(t-1)\}\text{.}
$
\subsection{Age of Information}
The \ac{AoI} represents how ``fresh" is the information from the perspective of the receiver. Let $A_{k}(t)$ be a strictly positive integer that depicts the \ac{AoI} associated with user $k$ at the receiver. If the received packet has been sampled during  slot $t$ and its transmission is successful, then $A_{k}(t+1)=1$. \ac{AoI} takes the value of one because the successful transmission takes one slot to be performed. On
the other hand, if the received packet has been sampled during the previous slots, then the age of information depends also in the time of the packet waiting for successful transmission. In this case, $A_{k}(t+1)> 1$. Therefore, the value of \ac{AoI} depends on the waiting time of the packet in the cache of the corresponding user. Furthermore, we assume that the value of AoI is bounded by an arbitrarily large finite value $M \in \mathbb{Z}_+$. This assumption is considered for two reasons: 
	1) In practical applications, values of \ac{AoI} that are larger than a threshold will not provide us additional  information about the staleness of the packet, \cite{ceran2019reinforcement,zhou2019joint}, 
	2) Assuming unbounded \ac{AoI} will  complicate  significantly our analysis without giving additional insights for the performance of the system.
\rev{Moreover, this assumption has been widely used in recent works that study the average \ac{AoI}, \cite{ceran2021reinforcement,maatouk2020optimality, hatami2021aoi}.}
Let $A_k^p(t)$ represent the system time of the packet in queue $k$, i.e, the waiting time of the packet. By definition, we have that 
$
	A_k^p(t) = t - \tau_k^s (t)
$, where $\tau_k^s(t)$ is the most recent sampling time. Naturally, the value of $\tau_k^s(t)$ changes only if  a new packet is sampled at the beginning of slot $t$. We consider that the decisions are taken at the beginning of each slot. 
In order to avoid having values of $A_k^p(t)$ that are greater than the \ac{AoI} at the destination, we bound the value of  $A_k^p(t)$ as 
$
	A_k^p(t) = \min\{t-\tau _{k}^s (t),M-1\}\text{, } \forall k\text{.}
$
Note that, when $A_k^p(t)$ reaches the value of $M-1$, the user drops the packet, and its cache becomes empty. 
We assume that a packet's transmission takes one time slot to be performed.
The evolution of the \ac{AoI} at the receiver for user $k$ is
\begin{align}\label{eq: EvolutionAoI1}
	A_k(t+1)=
	\begin{cases}
		A_k^p(t) + 1\text{, if } d_k(t+1)=1\text{,} \\
		\min\left\{A_k(t) + 1,M\right\}\text{, if } d_k(t+1) = 0\text{.}
	\end{cases}
\end{align} 
If the received packet has been sampled in the slot $t$, then $A_k^p(t)=0$  and therefore, $A_{k}(t+1)=1$. Alternatively, the evolution of \ac{AoI} can be written compactly as 
\begin{align}
	A_k(t+1) = d_k(t+1)(A_k^p(t)+1) 
+ (1-d_k(t+1)) \min\{A_k(t)+1,M\}\text{.}
	\label{eq: EvolutionAoI2}
\end{align}
\begin{figure}[t!]
	\centering
	\includegraphics[scale=0.3]{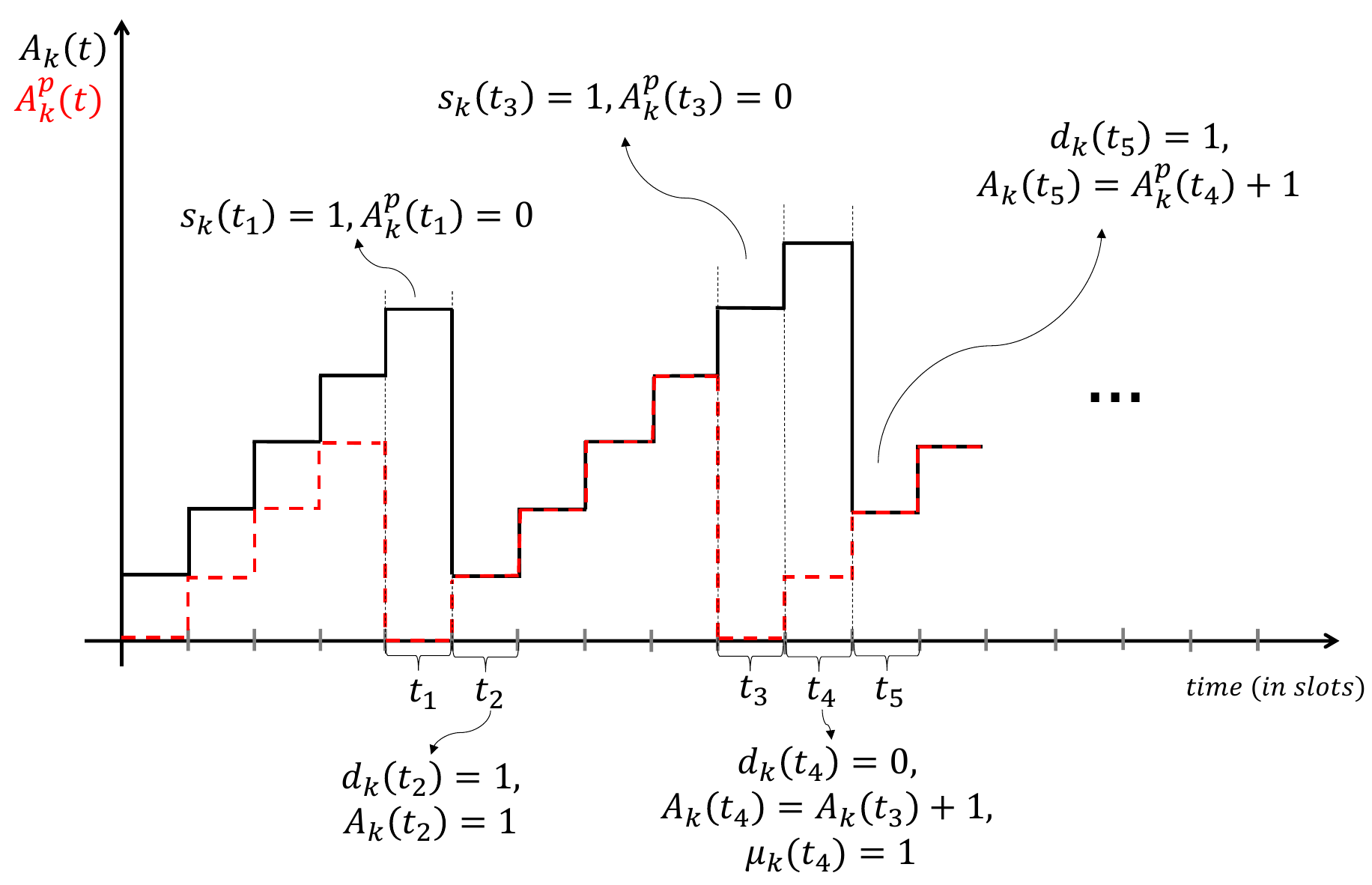}
	\caption{Example of the \ac{AoI} evolution.}
	\label{fig:age}
\end{figure}
Fig. \ref{fig:age} depicts an example of the evolution of  $A_k^{p}(t)$ and the \ac{AoI}  at the receiver. In the beginning of  time slot $t_1$, user $k$ is decided to sample and transmit fresh information. Therefore, $A_k^{p}(t)$  becomes  zero, and the packet is received successfully by the receiver after one slot. Therefore, in time slot $t_2$, the \ac{AoI} at the receiver is  one.  In time slot $t_3$, user $k$ is decided to sample and transmit a fresh packet, but the transmission fails. Thus, we observe by time slot $t_4$ the increase on \ac{AoI} in Fig. \ref{fig:age}.
 However, in time slot $t_4$, user $k$ is decided to transmit the old packet which is successfully received by the receiver after one slot.

For each transmission and sampling, we consider a corresponding cost. 
Let $c_s$ and $c_\text{tr}$ be the sampling and transmission cost, respectively. We consider that the costs remain the same over the time. The cost function for each user $k$ at each time slot $t$ is described as
$c_k(t) = \mu_k(t)c_{tr} + s_k(t) (c_s + c_{tr})\text{,}$
and the total system cost in time slot is described as
$c(t) = \sum\limits_{k=1}^K c_k (t)\text{.}$
The expected  average cost and the expected  average age for each user are defined as
\begin{align}
\bar{c} \triangleq \lim _{t \rightarrow \infty} \frac{1}{t} \sum_{\tau=0}^{t} \mathbb{E}\{c(\tau)\}\text{, }
\bar{A}_{k} \triangleq \lim _{t \rightarrow \infty} \frac{1}{t} \sum_{\tau=0}^{t} \mathbb{E}\left\{A_{k}(\tau)\right\}, \forall k \in \mathcal{K}\text{,}
\end{align}
respectively.

\subsection{Problem Formulation}
With the definitions of AoI and average costs, we define the stochastic optimization problem as following. 
\begin{subequations}\label{eq:optproblem1}
	\begin{align}
		\min\limits_{\bm{\mu}(t)\text{, } \mathbf{s}(t)} \quad & \bar{c}\\
		\text{s.~t.} 
		\quad & \bar{A}_{k}\leq A_{k}^{\text{max}}\text{, } \forall k \in \mathcal{K}\label{AoIconst}\text{,}\\
		 \quad & \sum\limits_{k=1}^{K} \mu_{k}(t) \leq 1 \text{, } \sum\limits_{k=1}^K s_{k}(t) \leq 1 \text{, } \label{InterConstr2	} \mu_k(t) + s_k(t) \leq 1\text{, } \forall k \in \mathcal{K}\text{,}\\
		\quad & \bm{s}(t)\text{, } \bm{\mu}(t) \in \{0,1\}^K\text{,}
	\end{align}
\end{subequations}
where $\bm{\mu}(t)= [\mu_{1}(t),\ldots,\mu_{K}(t)]$ and $\bm{s}(t) = [s_{1}(t),\ldots,s_{K}(t)]$.

Our goal is to find scheduling policies that minimize the total average cost while providing   average \ac{AoI} for each user $k$ below the value $A_{k}^{\text{max}}$.

\section{Scheduling Policies}
In this section, we provide three different scheduling policies that satisfy the average \ac{AoI} constraints and approximately minimize the total average cost. The first policy, named \acf{DPP}, is considered a fully-centralized policy that takes decision slot by slot in order to minimize the average cost of the system. It is derived by using tools from Lyapunov optimization theory \cite{neely2010stochastic}. The second policy, named \acf{OFRP}, is a stationary randomized policy where the scheduler schedules up to one user per time slot with some probability. Then, the scheduled user decides about its action. The third policy, named \acf{FoRP}, is a simple version of the second policy. The scheduler schedules up to one user per time slot with some probability. In this case, the scheduled user has only two options: either to sample and transmit or to remain silent.

Note that for the Lyapunov-based scheduling policy, the scheduler needs the complete information about the caches of the users. In addition, an optimization problem is solved slot by slot for  providing the optimal scheduler's decision. On the other hand, the stationary randomized policies do not require the scheduler to have any information about the caches of the users thus, it might be simpler to implement. Based on our analysis, we solve an optimization problem once, in the beginning, for optimizing the scheduling probabilities. These probabilities remain the same over the same, and there is no need for extra computation for every time slot.

\subsection{Drift-plus-penalty Policy}
In this section we provide a dynamic algorithm based on Lyapunov optimization theory that takes decisions slot by slot in order to minimize the total average cost while satisfying the time average \ac{AoI}.

In order to satisfy the  average constraints in \eqref{AoIconst}, we apply the methodology first developed  in \cite{neely2010stochastic}. In particular, each  average constraint in \eqref{AoIconst} is mapped into a virtual queue. We show that satisfying  the  average \ac{AoI} constraints  is equivalent to a queue stability problem. 

Let $\{X_k(t)\}_{k\in \mathcal{K}}$ be the virtual queues associated with constraints in  \eqref{AoIconst}. The evolution of each queue $k$ is shown below
\begin{align}\label{vqevolution}
	X_{k}(t+1) = \max [X_{k}(t)-A_{k}^{\text{max}}, 0] + A_{k}(t+1)\text{, } \forall k \in \mathcal{K}.
\end{align}
Process $X_{k}(t)$ can be viewed as a queue with ``arrivals" $A_{k}(t)$ and service rate $A_{k}^{\text{max}}$. 
\begin{lm}\label{lm:stability}
	If $X_k(t)$, is rate stable, $\forall k \in \mathcal{K}$, then the constraints in $\eqref{AoIconst}$ are satisfied.
\end{lm}
\begin{proof}
Using the basic sample property \cite[Lemma 2.1]{neely2010stochastic}, we have
	\begin{align}
			\frac{X_k(t)}{t} - \frac{X_k(0)}{t} \geq \frac{1}{t} \sum\limits_{\tau=0}^{t-1} A_k(\tau) - \frac{1}{t} \sum\limits_{\tau=0}^{t-1} A_k^{\text{max}}\text{, } \forall k \in \mathcal{K}.
	\end{align}
Therefore, if $X_k(t)$ is rate stable, so that $\frac{X_k(t)}{t}\rightarrow 0$,  then constraints \eqref{AoIconst} are satisfied with probability one \cite{neely2010queue}.

\end{proof}
Before describing the details of the analysis, let us recall a basic theorem  \cite{MeynMarkovChains}. Consider a system with $\mathcal{I} = \{1,2,\ldots,I\} $ queues. The number of unfinished jobs of queue $i$ is denoted by $q_{i}(t)$ and $\mathbf{q}(t) = \{q_{i}(t)\}_{i \in \mathcal{I}}$. The Lyapunov function and the the Lyapunov drift are denoted by $L(\mathbf{q}(t))$ and $\Delta(\mathbf{q}(t)) \triangleq \mathbb{E}\{L(\mathbf{q}(t+1))-L(\mathbf{q}(t))|\mathbf{q}(t)\}$, respectively.\\
\noindent \textit{Definition 1 (Lyapunov Function)}: A function $L: \mathbb{R}^{K} \rightarrow \mathbb{R}$ is said to be a Lyapunov function if it has the following properties:
1) It is non-decreasing in any of its arguments,
2) $L(\mathbf{x}) \geq 0\text{, } \forall \mathbf{x} \in \mathbb{R}^{K}$\text{,}
3)	 $L(\mathbf{x}) \rightarrow + \infty$, as $\|\mathbf{x}\| \rightarrow + \infty$.

\begin{thm}[Lyapunov Drift]  \label{thm:lyapunov}
If there are positive values B, $\epsilon$ such that for all time slots $t$ we have $\Delta(\mathbf{q}(t)) \leq B - \epsilon \sum\limits_{i=1}^I q_i (t)$, then the system $\mathbf{q}(t)$ is strongly stable \cite{MeynMarkovChains}.
\end{thm}
The intuition behind Theorem \ref{thm:lyapunov} is that if we have a queueuing system and   we provide a policy for which the Lyapunov drift is bounded by a constant value $B>0$ and the sum of the length of the queues multiplied by a negative value for every time slot, then the system can be stabilized. 

The \ac{DPP} algorithm is designed to minimize the sum of the Lyapunov drift and a penalty function \cite[Chapter 3]{neely2010stochastic}. First, we define the Lyapunov drift as
\begin{align}\label{eq:drift}
	\Delta (\mathbf{X}(t)) = \mathbb{E} \{ L(\mathbf{X}(t+1))  - L(\mathbf{X}(t))  | S_t\} \text{,}
\end{align}
where $S_t = \{A_k(t), X_k(t)\}_{k\in\mathcal{K}}$ is the network state at  slot $t$, and $\mathbf{X}(t) = 
\{X_{k}(t)\}_{k\in \mathcal{K}}$. The associated Lyapunov function is defined as 
$  
	L (\mathbf{X}(t))= \frac{1}{2} \sum\limits_{k=1}^{K} X^2_{k}(t). 
$
The above expectations are with respect to the channel randomness and the scheduling policy. We apply the \ac{DPP} algorithm in order to minimize the total average cost (penalty function) while stabilizing the virtual queues, i.e., providing  average \ac{AoI} below the given value  for each user. In particular, this approach seeks to minimize an upper bound of the following expression 
\begin{align}\label{expr: driftpluspenalty}
	\Delta (\mathbf{X}(t))  + V \mathbb{E}\{c(t) |S_{t}\}\text{,}
\end{align}
where $V$ is an importance weight factor. An upper bound for the expression in \eqref{expr: driftpluspenalty} is shown below 
\begin{align}\nonumber
	\Delta(\mathbf{X}(t)) + V\mathbb{E}\{c(t)|S_t\}  
	&\leq B + \sum\limits_{k=1}^{K} X_k(t) \left[\right.\mathbb{E}\{W_k(t)(A_k^p(t)+1)  + (1-W_k(t))\times\\
	& \min\left[A_{k}(t)+1, M\right]| S_t\}
	 -A_k^{\text{max}} \left.\right]   + V\mathbb{E}\{c(t)|S_t\}\text{,}
	 	 \label{eq: DriftBound}
\end{align}
where $B\geq \sum\limits_{k=1}^K \frac{\mathbb{E}\{A_k^2(t+1)|S_t\} +(A_k^{\text{max}})^2 }{2}$, and $W_{k}(t) = p_ks_k(t)+p_k\mu_k(t)$. The complete derivation of the above bound can be found in Appendix A. If we set $B = \sum\limits_{k=1}^K  \frac{M^2 + (A_{k}^{\text{max}})^2}{2}\geq  \sum\limits_{k=1}^K \frac{\mathbb{E}\{A_k^2(t+1)|S_t\} +(A_k^{\text{max}})^2 }{2}$, we see that $B$ is a constant and it does not depend on the scheduling decisions over slots. Therefore, we can exclude $B$ from the optimization problem. The \ac{DPP} algorithm takes sampling and transmission decisions at each time slot by solving the following optimization problem. 

\begin{subequations}\label{OptDPP}
	\begin{align}\nonumber
		\min\limits_{\bm{\mu}(t)\text{, }\bm{s}(t)} \quad &   \sum\limits_{k=1}^K  \{
		X_{k}(t) [(A_k^p(t)+1 )W_{k}(t) +\\ \nonumber
		& \min\{ (A_{k}(t)+1), M\} (1-W_{k}(t)) - A^\text{max}_k] \} \\
		& + Vc(t)\\
		\text{s.~t.} \quad & \sum\limits_{k=1}^{K} \mu_{k}(t)\leq 1\text{, }
		\sum \limits_{k=1}^K s_k(t) \leq 1\text{,} \mu_k(t) + s_k(t) \leq 1\text{, } \forall k \in \mathcal{K}\text{,}\\
		\quad & \bm{s}(t)\text{, } \bm{\mu}(t) \in \{0,1\}^{K} \text{.}
	\end{align}
\end{subequations}
Also, we define  function $y_k(t+1)$, $\forall k$ as following
\begin{align}\nonumber
	& y_k(t+1)  = A_k(t+1) - A_k^{\text{max}} \\\nonumber
	& =\mathbb{E}	\{(A_{k}^p(t)+1)(p_{k}s_{k}(t)+p_{k}\mu_{k}(t)) 
	+ \min\{(A_{k}(t)+1),M\}(1-p_ks_k(t)-\mu_{k}(t)p_k\} 
	- A_k^{\text{max}} \text{.}
\end{align}

\rev{
\begin{lm}\label{Lm}
	We consider a class of stationary possible randomized policies denoted by $\Omega$. A policy $\omega(t)$ that belongs to the class $\Omega$ is an i.i.d. process that takes probabilistc decisions independently of the states of the caches at every time slot $t$. Let $y_k(t)=A_k(t)-A_k^{\text{max}}$, and $c(t)$ be the total system cost. Then, if the problem in \eqref{eq:optproblem1} is strictly feasible, and the second moments of $y_k(t)$ and $c(t)$ are bounded, then for any $\epsilon>0$, there is an $\omega(t)$ policy under which the following holds:
	\begin{align}\label{eq:costbound}
		 \mathbb{E} \{y^*_k(t+1)\} \leq \epsilon\text{, }
		\mathbb{E} \{c^*(t)\} = c_{\omega} (\epsilon) \leq c^{\text{opt}} + \epsilon\text{,}
	\end{align}
	where $y_k^{*}(t+1)$ and $c^*(t)$ are the resulting values of the $\omega$ policy, $c^{\text{opt}}$ is best objective function in \eqref{eq:optproblem1} achievable by any stationary randomized policy, and $ c_{\omega}(\epsilon) $ is a feasible suboptimal solution to the problem in \eqref{eq:optproblem1} that can be achieved by an $\omega$ stationary randomized policy. 
\end{lm}
\begin{proof}
	We consider the cases in which the problem in \eqref{eq:optproblem1} is strictly feasible. Furthermore, by definition, $A_k(t)$, $\forall k$, is bounded by finite value $M$, therefore, the first and the second moments of $A_k(t)$, are also bounded. Furthermore, the cost function $c(t)$ is also bounded by the sum of constant values $c_\text{s}+c_{\text{tr}}$, i.e., the sampling and the transmission cost, respectively. Therefore, we have:$
		1 \leq \mathbb{E}\left\{A^2_k(t)\right\} \leq M^2 \text{, } 
		0  \leq \mathbb{E}\left\{c^2(t)\right\}  \leq (c_{\text{tr}} + c_\text{s})^2K^2\text{.}$
	Then, the boundness assumptions in \cite[Ch. 4.2.1]{neely2010stochastic} are satisfied. Therefore, from Theorem 4.5 in \cite{neely2010stochastic}, we get the result.
\end{proof}
}

\rev{
\begin{thm}\label{Thm:Feasibility}
	The \ac{DPP} algorithm satisfies any feasible set of the maximum \ac{AoI} constraints.
\end{thm}
\begin{proof}
	See Appendix B.
\end{proof}
}

\rev{
\begin{thm}[Optimality of the DPP algorithm]
  After applying the DPP algorithm, we obtain an expected average cost that  is bounded as
\begin{align}\label{bound: cost}
	\lim_{t\rightarrow \infty} \sup \frac{1}{t} \sum\limits_{\tau=0}^{t-1} \mathbb{E} \{c(\tau)\} \leq c^{\text{opt}} +\frac{B}{V}\text{.}
\end{align}
\end{thm}
\begin{proof}
	See Appendix C.
\end{proof}
}
\textbf{Remark 1. }Theorem 2  indicates that the DPP algorithm provides a solution arbitrarily close to the optimal. We can get better performance in terms of  average cost by increasing the value of $V$. However, by increasing the value of $V$, the values of the length of the virtual queues will increase as well. Therefore, there is a trade-off between the   average cost and  average age.

\rev{\textbf{Remark 2. }The \textit{drift-plus-penalty} algorithm is based on the Lyapunov optimization theory. The Lyapunov drift is used to guarantee the stability of the virtual queues, and therefore, the satisfaction of the time average constraints. The Lyapunov drift is the expected change in the values of $\mathbf{X}(t)$ from one slot to another. By providing an upper bound on the Lyapunov drift (eq. (19)), we guarantee that minimizing this bound that is finite, the length of the virtual queues are also bounded and therefore stable. The \textit{drift-plus-penalty} seeks to minimize the bound on the Lyapunov drift and the total system cost.}

\rev{
We prove that, if we solve the weighted optimization problem that minimizes the bound of the Lyapunov drift plus the penalty, we achieve asymptotically the  minimum cost and we satisfy the constraints. The weighted optimization problem consists of two kinds of weights. The first is the importance factor $V$ multiplied by the total system cost. The second kind of weight is the values of the virtual queues. Therefore, when the length of a virtual queue is quite large and larger than the corresponding cost, we allocate resources to the corresponding user in order to reduce the \ac{AoI} at the receiver. On the other hand, if the length of a virtual queue is low, then implicitly we know, that the corresponding user had been allocated a sufficient amount of resources so far, such that its \ac{AoI} at the receiver is low.}
\subsection{First Stationary Randomized Policy - Fresh or Old Randomized Policy}
We consider a  scheduler that decides probabilistically which user will be scheduled at every time slot. The scheduler schedules up to one user per time slot. The probability of the scheduler allocating user $k$ is $\alpha_k$.  Therefore, $\sum\limits_{k=1}^K\alpha_k = 1$. If a user $k$ is scheduled in a slot, then it decides by itself the following actions with the corresponding probabilities:
\begin{itemize}
	\item If there is a packet in the cache, user $k$:
		1) samples and transmits \ac{w.p.} $u_k$,
		2) transmits an old packet  \ac{w.p.} $q_k$,
		3)  remains silent \ac{w.p.} $1 -u_k-q_k$.
	\item If there is not packet in the cache, user $k$
		1) samples and transmits \ac{w.p.} $u_k^{\prime}$,
		2) remains silent \ac{w.p.} $1-u_k'$.
\end{itemize}
Note that
$
	\Pr\left\{s_{k}(t)  = 1|Q_k(t)=1\right\}  =   \alpha_k u_{k}\text{,}
	 \Pr\left\{s_{k}(t)   = 1|Q_k(t)=0\right\}  =   \alpha_ku_{k}'\text{.}
$
Furthermore, 
$
	\Pr\left\{\mu_{k}(t) = 1 | Q_k(t) = 1\right\}= \alpha_kq_k\text{.}
$
 In the case where $A_k(t)= A_k^p(t)+1$, there is no advantage regarding the \ac{AoI} to transmit an old packet because after one slot the \ac{AoI} of the received packet will be $A_k(t+1)=A_k^p(t)+1=A_k(t)$. Therefore, in this case, we discard the packet in the cache, and then the options are either to sample and transmit a fresh packet   or to remain silent.

In order to select a set of decision probabilities that satisfies the constraints while minimizing the average cost, we need to derive the average cost and the average \ac{AoI} as function of those probabilities. Here, we focus on the case of a single user. However, the same methodology is applied for each user independently in a multiple-user scenario. For the sake of the presentation, we omit the subscript $k$ without sacrificing clarity. In order to obtain the expression for the distribution of the \ac{AoI} and the average \ac{AoI}, we model the evolution of the \ac{AoI} at the receiver, $A_t=A(t)$, and the evolution of the waiting time in the cache, $A_t^p=A^p(t-1)+1$, as one two dimensional \ac{DTMC}.  The transition to the next state depends on the current state, i.e., $(A_t^p, A_t)$, on the decision of the scheduler, and on the decision of the scheduled user.
The \ac{DTMC}  $\{(A_t^p, A_{t})\}$ is described by the following transition probability, $\forall i,j,m,l,
		P_{(i,j)\rightarrow (m,l)} = \Pr\left\{ A^p_{t+1} = m, A_{t+1}=l\text{ }|\text{ }A_t^p=i, A_t = j \right\}\text{.}$
Note that we do not need to include the state of the cache in the Markov chain because the values of $A^p_{t}$ and $A_t$ can indicate us whether the cache has a packet or not. For example, if $A_t^p=A_t$, we know that in the begining of time slot $t$, we have a successful transmission. Therefore, the cache of the user is empty after the successful transmission and the only options are either to sample and transmit or remain idle.  The transition to each state depends on the events happened in the previous slot. The events are: i) the decision of the scheduler, ii) the decision of the scheduled user, if  there is any, iii) the outcome of the transmission. We categorize the states according to the values of $A_t^p$ and $A_t$ below.\\
\noindent \underline{If $i\leq j-1$, $j< M$, $i<M-1$}:
In this case, there is a packet in the cache sampled during the previous slots, i.e., $Q(t)=1$. The transition probabilities are:
\begin{align}
	P_{(i,j)\rightarrow (m,l)}  = 
	\begin{cases}
		\alpha u p\text{ if, } m=1 \text{ and } l=1\text{, }\\
		\alpha u (1-p)\text{, if } m=1\text{ and } l=j+1\text{, } \\
		\alpha q p \text{, if } m=i+1 \text{ and } l=i+1\text{, }\\
		1  - \alpha u - qp\alpha\text{, if } m=i+1 \text{, } l=j+1\text{.}
	\end{cases}
\end{align}

\noindent \underline{If  $j<M$, $i<M-1$:}
In this case, there is no packet in the cache, i.e., $Q(t)=0$. The transition probabilities are described below:
\begin{align}
	P_{(i,j)\rightarrow (m,l)} = 
	\begin{cases}
		\alpha u' p\text{, if } m=1\text{, } l=1\text{,}\\
		\alpha u'(1-p) \text{, if } m=1\text{, }l=j+1\text{,}\\
		1 - \alpha u' \text{, if } m=i+1\text{, }l=j+1 \text{.} 
	\end{cases}
\end{align}

\noindent \underline{If  $j=M$, $i<M-1$:}
In this case, we have a packet in the cache. 
\begin{align}
	P_{(i,j)\rightarrow (m,l)} = 
	\begin{cases}
		\alpha u p\text{, if } m=1\text{, } l=1\text{,}\\
		\alpha u(1-p) \text{, if } m=1\text{, }l=M\text{,}\\
		\alpha q p \text{, if } m=i+1\text{, } l=i+1\text{,}\\
		1 - \alpha u -qp\alpha \text{, if } m=i+1\text{, }l=M \text{.} 
	\end{cases}
\end{align}

\noindent\underline{If  $i=1$, $j=M$.}
In this case, there is an old packet packet in the cache, $Q(t)=1$.
The transition probabilities are described below:
\begin{align}
	P_{(i,j)\rightarrow (m,l)}=
	\begin{cases}
		\alpha u p\text{, if } m=1 \text{ and } l=1\text{,}\\
		\alpha u (1-p)\text{, if  } m=1 \text{ and } l=M\text{,}\\
		\alpha qp \text{, if } m=2 \text{ and } l=2\text{,}\\
		1-\alpha u - \alpha qp\text{ if } m=i+1 \text{ and } l=M\text{.}
	\end{cases}
\end{align}



\vspace{10mm}
\noindent \underline{\ac{AoI} in the boundaries: $i=M-1$, $j=M$:}
In this case, even if the cache has an old packet the user drops this packet because a successful transmission of the old packet will not improve the AoI at the receiver. The transition probabilities are described below:
\begin{align}
		P_{(i,j)\rightarrow (m,l)}=
	\begin{cases}
		1-\alpha u' \text{, if } m=M-1\text{, } l=m\text{,}\\ 
		 \alpha u' p\text{, if  } m=1 \text{, } l=1\text{,}\\
		 \alpha u' (1-p)\text{, if } m=1\text{, } l=M\text{.}
	\end{cases}
\end{align}

We have now completely describe the transition matrix, $\bm{P}$, and we can obtain the steady steady distribution of AoI at the receiver. We consider one Markov Chain for each
user $k$ and its transition matrix is denoted by $\bm{P}_{(k)}$.

We denote the steady state distribution of the AoI by a row vector
\begin{align}
	\bm{\pi}_{(k)}=\left[\pi_{0,1}^{(k)}, \pi_{0,2}^{(k)}, \ldots, \pi_{0, M}^{(k)}, \pi_{1,2}^{(k)}, \ldots \pi_{1, M}^{(k)}, \ldots\right]\text{.}
	\label{eq:steadystate}
\end{align}
We obtain $\bm{\pi}_{(k)}$ by solving the following linear system of equations
$
	\begin{aligned} \bm{\pi}_{(k)} \boldsymbol{P}_{(k)} =\boldsymbol{\pi}_{(k)}\text{, }
	 \boldsymbol{\pi}_{(k)} \mathbf{1} =1 \text{,}
	\end{aligned}
$
where $\mathbf{1}$ is a column vector with all its elements being one. 
We can obtain the steady state distribution by applying numerical methods, e.g., \ac{EVD}. The average AoI for each user $k$  is calculated as
		$	\bar{A}_k = \sum\limits_{i=0}^{M-1} \sum\limits_{j=i+1}^{M} j\pi_{i,j}^{(k)} \text{.}$
We set $\theta = \sum\limits_{i=1}^{M-1}\sum\limits_{j=i}^{i+1} \pi^{(k)}_{i,j}$, where $\theta$ is the probability of the cache being empty.
The average cost for each user $k$ is calculated as
\begin{align}
	\bar{c}_k = \theta \alpha_k u_k'(c_{tr} + c_s) + (1-\theta) (\alpha_k(q_kc_{tr} + u_k(c_{tr} +c_s) )) \label{eq:avercost}\text{.}
\end{align}
In order to optimize the parameters $\alpha_k$, $u^\prime_k$, $q_k$, we select a step for selecting combinations of the possible values of the probabilities. We discretize the space of the probabilities and we take the corresponding combinations. The smallest the step the highest the number of combinations and therefore, the complexity. We find the combinations for which the constraints are satisfied, and we select the combination of the probabilities that provide the lowest total system cost.

\subsection{Fresh-only Randomized Policy}

Here, we propose a simple randomized policy for which we obtain closed-form expressions.
At each time slot, the scheduler probabilistically schedules up to one user at each time slot. Then, the scheduled user decides either to sample and transmit or to remain idle (no option for  transmission of an old packet). In particular, the scheduler schedules user $k$ with probability $\alpha_k'$ where $\sum\limits_{k=1}^K \alpha_k'=1$. If user $k$ is scheduled, then
1) it decides to sample and transmit a fresh packet with probability $\phi_k$, 2) it decides to remain silent with $1 - \phi_k$.
In order to obtain the expression for the average \ac{AoI} for each user $k$, and the total average cost we model the system as one \ac{DTMC}. In the following analysis, we omit subscript $k$ without sacrificing clarity. The same methodology is applied for each user $k$. 
We model the AoI for each user $k$ as one Markov chain, where $\delta=\alpha' \phi p$. The transition matrix, $\mathbf{P'}$, of the Markov chain is shown below
\begin{small}
\begin{align}\nonumber
\mathbf{P'}&=
\left[\begin{array}{cccccc}
\delta & 1 - \delta & 0 & 0  & \cdots & 0\\
\delta & 	0     & 1 - \delta  & 0  & \cdots & 0 \\
\vdots &  	\vdots	  & 				 &  \ddots       \\ 
 \delta & 	0	   &      0 & 0 &			\cdots	  & 	1- \delta   \\
 \delta & 	0	   &      0 & 0 &			\cdots	  & 	1- \delta   
\end{array}    
\right]\text{.}
\end{align}
\end{small}
We consider one Markov chain for each user $k$. The transition matrix of each user $k$ is denoted by $\bm{P}'_{(k)}$.
We denote the steady state distribution of the \ac{AoI} by a row vector as
$
	\bm{\pi}_{(k)}' = \left[\pi_{1}^{(k)}, \pi_{2}^{(k)}, \ldots, \pi_{M}^{(k)}\right]\text{.}
$
In order to obtain the steady state distribution of the \ac{AoI}, we solve the following linear system of equations
$
		\bm{\pi}'_{(k)} \boldsymbol{P}'_{(k)} =\boldsymbol{\pi}_{(k)}'\text{, }
			 \boldsymbol{\pi}_{(k)}' \mathbf{1} =1 \text{.}
$
After some calculations, we obtain the steady state distribution of the AoI at the receiver, as follows:
$
			\pi_i^{(k)} =\delta_k(1-\delta_k)^{i-1}\text{, for } i<M
			\pi_M^{(k)}  = (1-\delta_k)^{M-1}\text{, for } i=M\text{.}
$
We set $\bar{\delta}_k = 1-\delta_k$, and the average \ac{AoI} for each user $k$ is calculated as
 \begin{align}\nonumber
   \bar{A}_k & = \sum\limits_{i=1}^{M} i \pi_i^{(k)} = \sum\limits_{i=1}^{M-1} i\delta_k\bar{\delta}_k^{i-1}  + M(\bar{\delta}_k)^{M-1}
   =  \frac{\delta_k}{\bar{\delta}_k}\sum\limits_{i=1}^{M-1} i\bar{\delta}_k^{i}+M(\bar{\delta}_k)^{M-1}\\
   &\stackrel{(a)}{=}\frac{(M-1)\bar{\delta_k}^M - M\bar{\delta_k}^{M-1}+1}{\delta_k} + M\bar{\delta_k}^{M-1}\text{,}
 \end{align}
 where $(a)$ follows by applying $\sum_{i=0}^{n} i c^{i}=\frac{n c^{n+2}-(n+1) c^{n+1}+c}{(c-1)^{2}}\text{, } c\neq 1$.
 The average cost for each user $k$ is 
$\bar{c}_k = (c_{tr} + c_s) (\alpha'_k \phi_k)\text{.}$
 
In order to minimize the probabilities, we calculate the average AoI for every combination of value of $\phi_k$ and $\alpha_k^\prime$ between $0$ and $1$ with a step $\Delta$, and if the consraint is satisfied, we calculate the the corresponding cost. The decision probabilities for every users are the ones for which the average \ac{AoI} constraints are satiesfied and the cost is minimized. 

\section{Simulation and Numerical Results}
In this section, we provide results that show the performance of the proposed policies regarding the total average cost, for various scenarios. We compare the scheduling policies for different  success probabilities, as well as,  different \ac{AoI} constraints. Furthermore, we analyze the scheduling decisions for the \ac{DPP} and OFRP policy for different values of costs, i.e., when the sampling cost is larger than the transmission cost.

In order to find the optimal probabilities of the randomized policies, we apply exhaustive search. In particular, we take values of the probabilities that are between $0$ and $1$, with step that equals $0.01$. We set  these values  in the derived expressions and we find the combinations for which the \ac{AoI} constraints are satisfied and the total average cost is minimized. For the \ac{DPP} policy, each simulation has run for $10^6$ slots, and the importance factor, $V$, is equal to $800$, for each case. We perform our simulations in MATLAB.

\subsection{Performance of  the Scheduling Policies}
\begin{figure}[t!]
	\centering
	\begin{subfigure}{0.45\textwidth}\centering\includegraphics[scale=0.4]{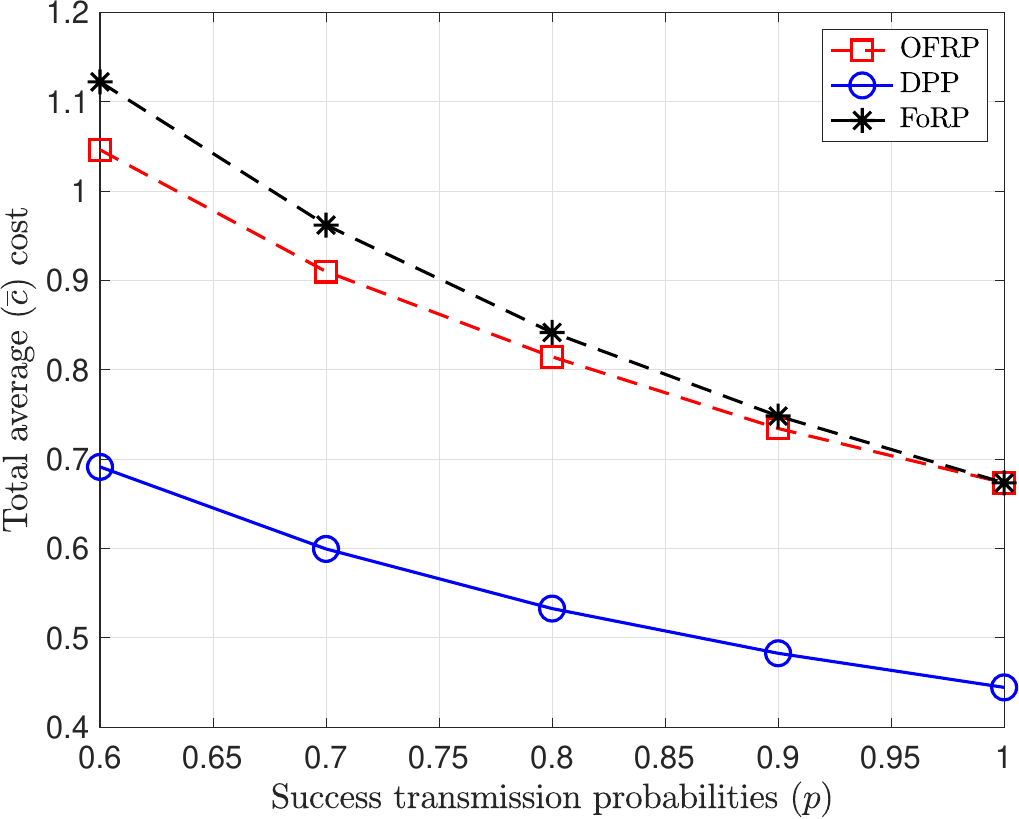}\caption{$A=5$, $M=10$.}\label{Fig:AvCostDiffSuccProbA5M10}\end{subfigure}
	\begin{subfigure}{0.45\textwidth}\centering\includegraphics[scale=0.4]{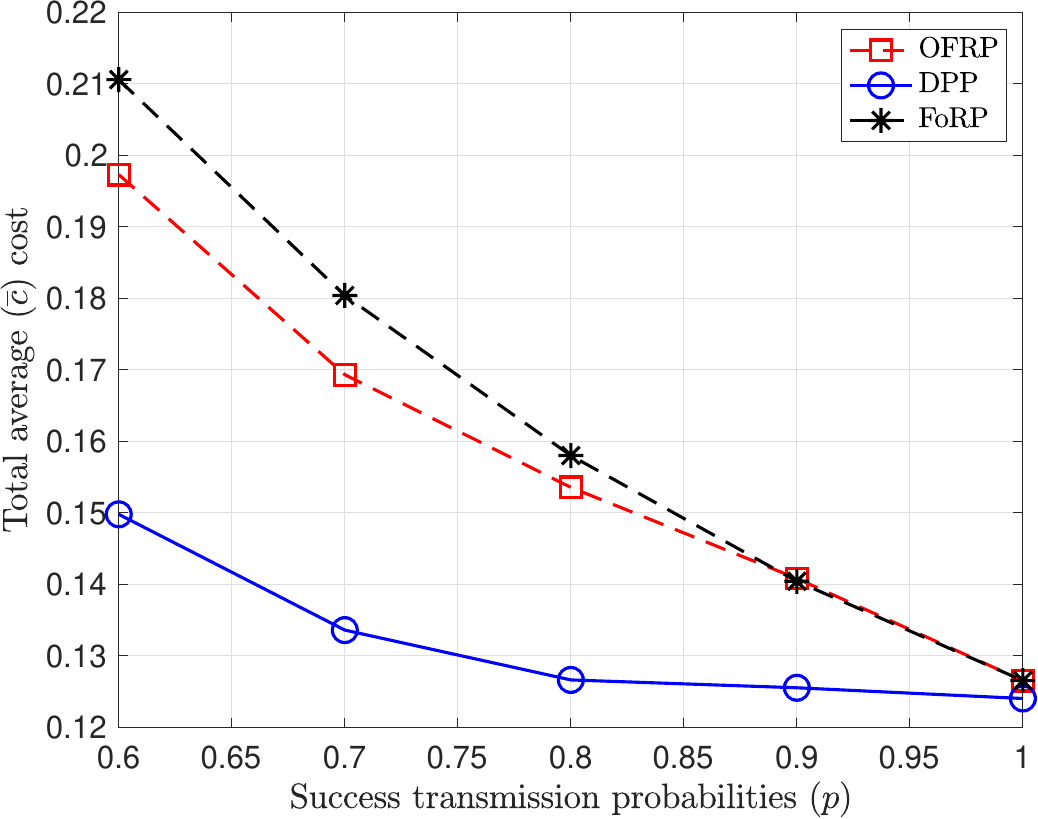}\caption{$A=15$, $M=20$.}\label{Fig:AvCostDiffSuccProbA15M20}\end{subfigure}
	\caption{Performance comparison between the proposed policies for different success probabilities.}
	\label{Fig: PerfComparison}
\end{figure}
In Fig. \ref{Fig: PerfComparison}, we provide results for different values of the success transmission probabilities, and different values of the \ac{AoI} constraints. In particular,  Fig. \ref{Fig:AvCostDiffSuccProbA5M10} depicts the total average cost for each scheduling policy, where the values of the \ac{AoI} constraints are $A=5$, and the \ac{AoI} threshold, $M$, is equal to $10$. We observe that when the success probabilities approach $1$, the performance of  \ac{OFRP} and \ac{FoRP} tend to be identical, and for $p=1$ the performance of the randomized policies is equal. The reason is that  for high success probabilities, the probability of the cache having a packet is quite small, and for error-free channels, the caches of the users are always empty. Therefore, \ac{OFRP} schedules always the users to sample and transmit or to remain silent. In this case, the two randomized policies have identical behavior. 

On the other hand, when the success transmission probabilities are small and $A =15$, the difference between  the randomized policies increases regarding the performance, as shown in Fig. \ref{Fig:AvCostDiffSuccProbA15M20}. In this case, the scheduling policies utilize the flexibility that is given by the large value of $A_1^{\text{max}}$. Therefore, in this case, the option of transmitting an old packet reduces the total average cost.

\begin{figure}
		\begin{subfigure}{0.5\textwidth}\centering\includegraphics[scale=0.4]{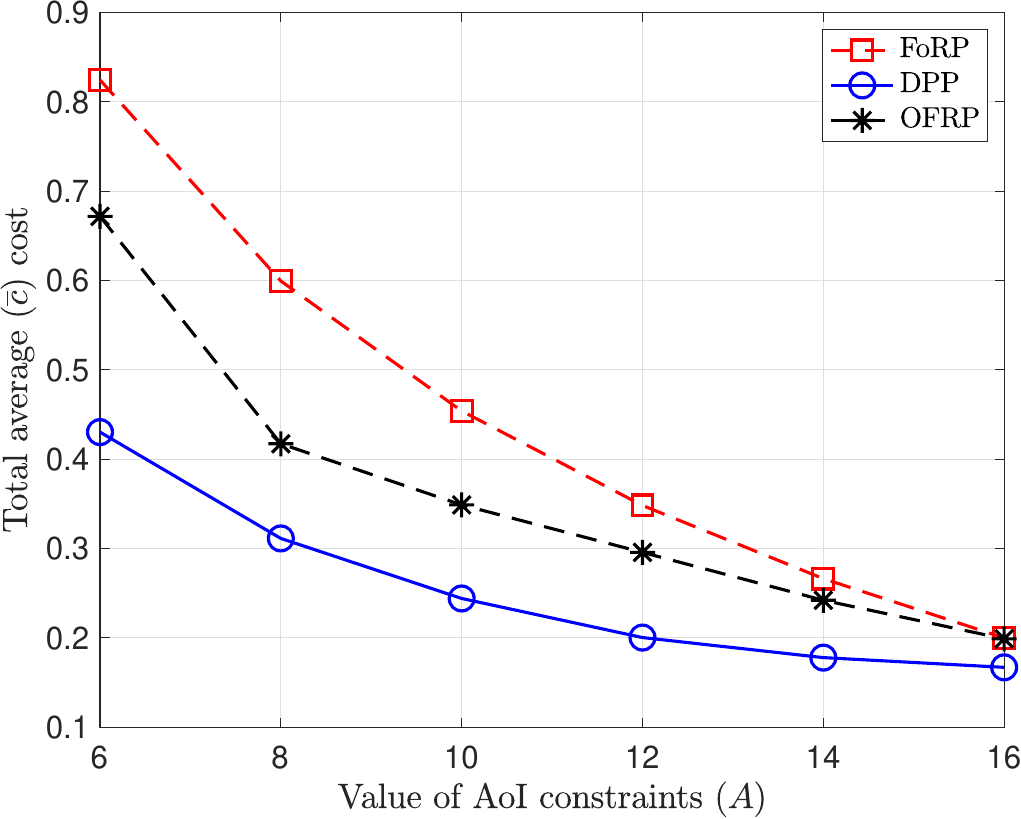}\caption{Total average cost.}\label{Fig:HardnessAoI}\end{subfigure}
	\begin{subfigure}{0.46\textwidth}\centering\includegraphics[scale=0.4]{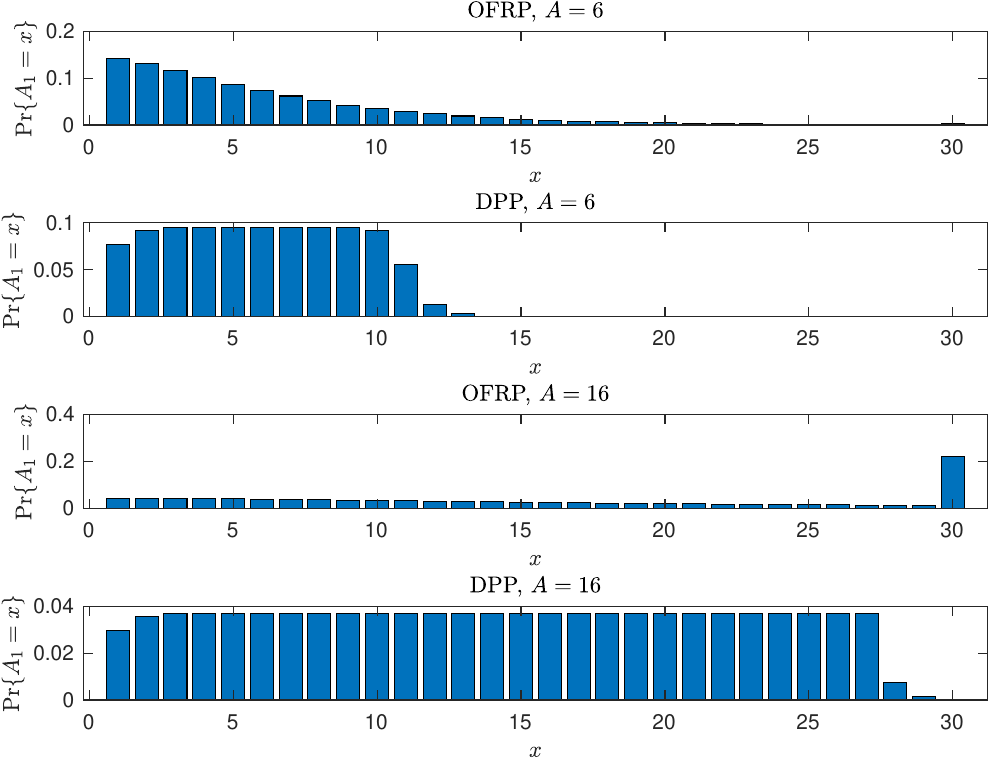}\caption{Distribution of the AoI of user $1$.}\label{Fig:Distribution}\end{subfigure}
	\caption{Performance comparison between the proposed policies for values of the \ac{AoI} constraints. $p=0.8$, $M=25$.}
\end{figure}
In Fig. \ref{Fig:HardnessAoI}, we provide results that show the performance of the scheduling policies as the values of the \ac{AoI} constraints, i.e., $A$, increases. We consider that the success transmission probabilities are equal to $0.8$, and the \ac{AoI} threshold, $M$, is $25$. Obviously, for larger values of the \ac{AoI} constraints the total average cost decreases, for all the scheduling policies. Furthermore, we observe that as the hardness of the \ac{AoI} constraints decreases, the difference in the performance between \ac{OFRP} and \ac{DPP} decreases as well. In this case, the \ac{OFRP} utilizes that the \ac{AoI} cannot be larger a value, and therefore, it gains in cost. On the other hand, we observe that the \ac{DPP} policy allows the value of \ac{AoI} to reach the threshold only for a small percentage of time (see the distribution of the \ac{AoI} in Fig. \ref{Fig:Distribution}, for $A_1^{\text{max}}=16$). However, the \ac{DPP} algorithm still outperforms the randomized policies regarding the total average cost.

\begin{figure}
	\centering
	\includegraphics[scale=0.41]{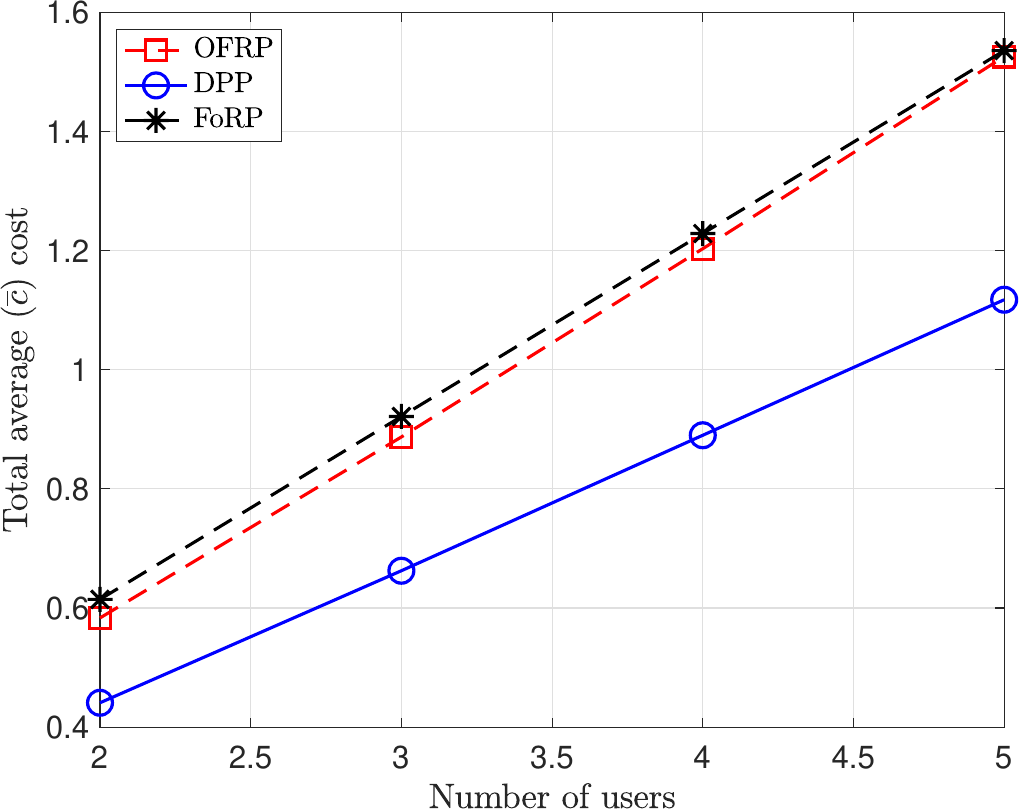}
	\caption{A multi-user scenario. $M=15$, value of AoI constraints: $A_1^{\max} = A_2^{\max} = 7$. Success transmission probabilities: $p_1=p_2=0.8$.}
	\label{fig:multiusers}
\end{figure}

\begin{figure}
	\centering
	\includegraphics[scale=0.4]{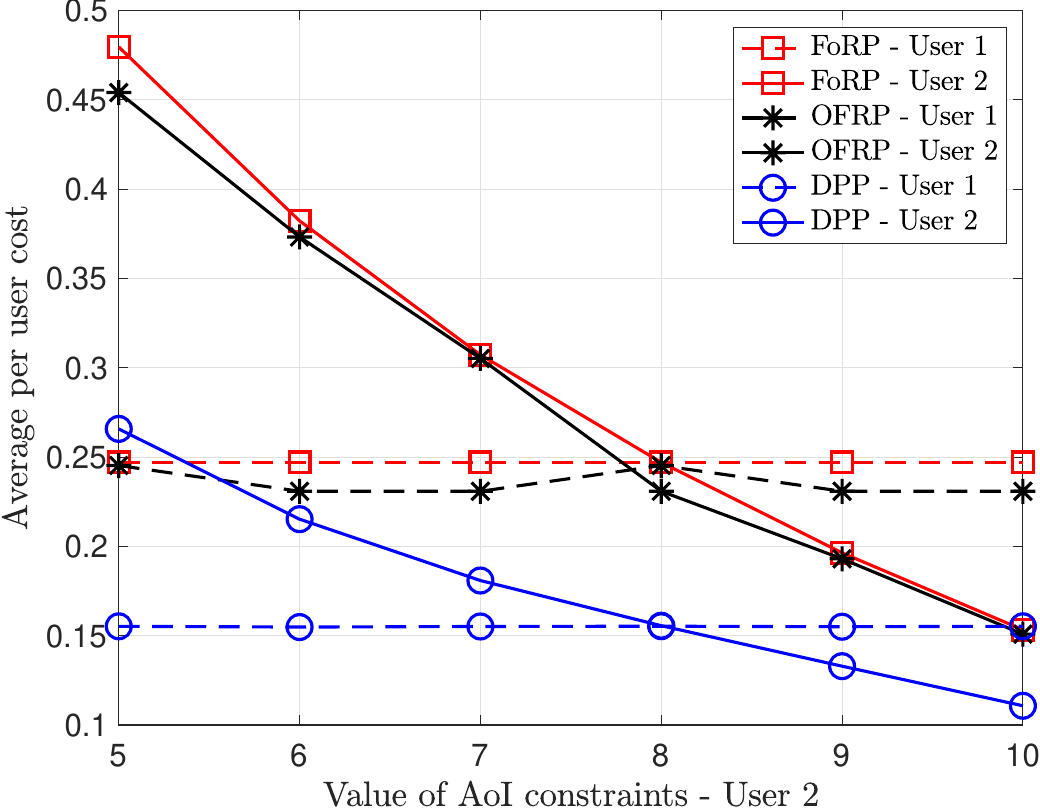}
	\caption{$p_1=p_2=0.8$. $M=20$. $A_1^{\max} = 8$.}
	\label{fig:figuresdifferentvaoinonsymmetric}
\end{figure}

In addition, we provide results for  multi-user scenarios in Fig. \ref{fig:multiusers}. The success transmission probabilities are equal to $0.8$, and the values of AoI constraints are equal to $7$. We observe that the total average cost increases linearly with the respect of the number of users in the system. However, the total average cost achieved by applying stationary randomized policies increase faster than the one achieved by the DPP algorithm. Therefore, the DPP algorithm can be considered more robust in terms of the total average with respect to the number of users in the system. Furthermore, in Fig. \ref{fig:figuresdifferentvaoinonsymmetric}, we provide results for two non-symmetric users. In particular, the value of \ac{AoI} constraint of user $1$ is equal to $8$.  We provide results for the average per user cost as the value of \ac{AoI} constraint of user $2$ increases. We observe that as the value of \ac{AoI} increases, the average per user cost decreases almost linearly for user $2$. On the other hand, the cost for user $1$ remains almost constant since the value of \ac{AoI} does not change. 



\subsection{Scheduling Decisions and Performance Comparison}
\begin{figure}[t!]
	\centering
	\begin{subfigure}{0.45\textwidth}\centering\includegraphics[scale=0.4]{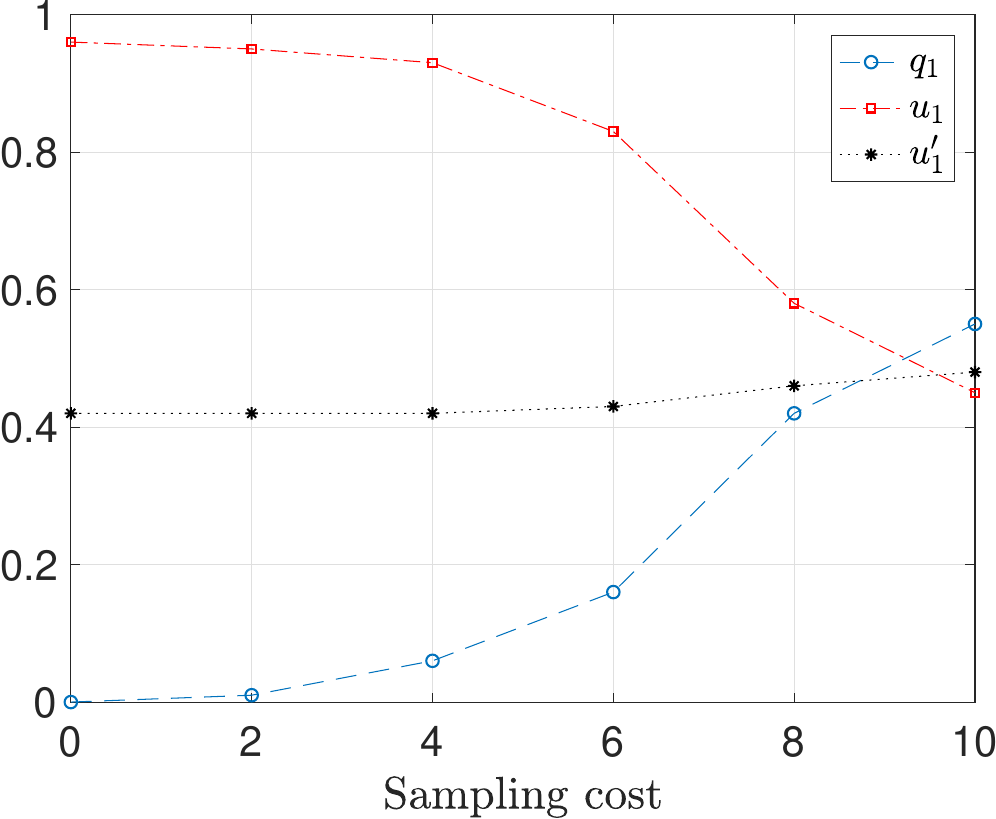}\caption{Optimized probabilities of the \ac{OFRP} policy.}\label{Fig:OptProbabilitiesDiffSCost}\end{subfigure}
	\begin{subfigure}{0.46\textwidth}\centering\includegraphics[scale=0.4]{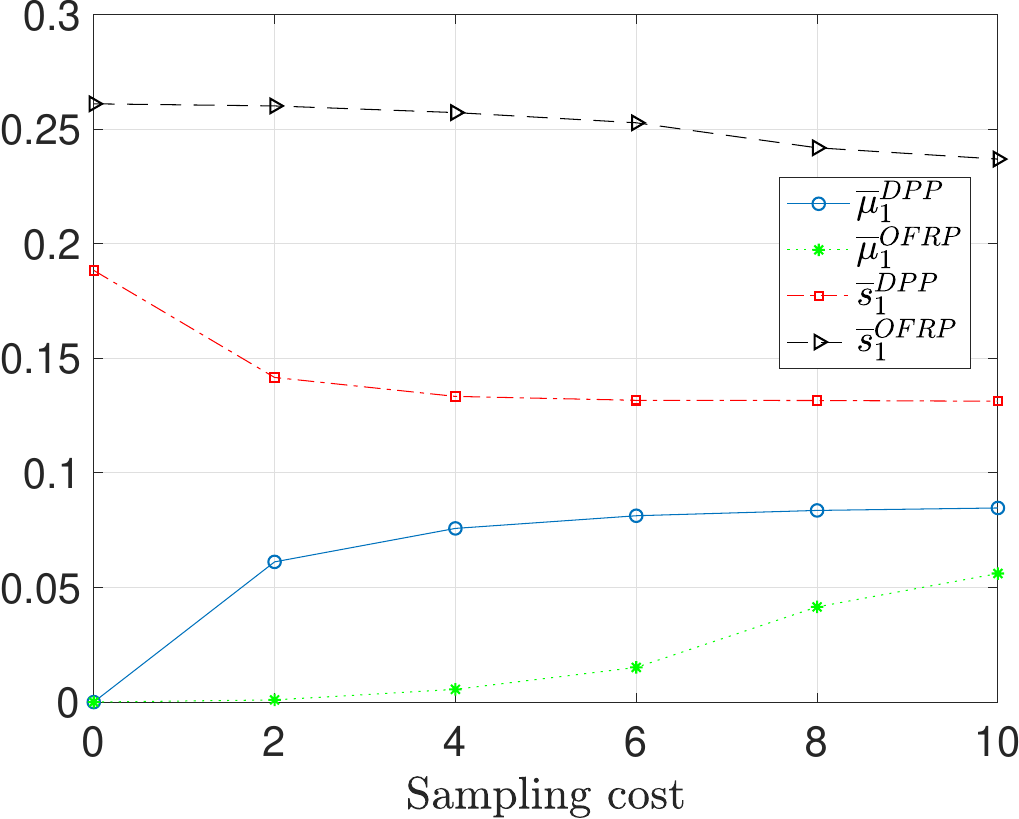}\caption{Average values of $\mu_1$ and $s_1$.}\label{Fig:AvValuesS1mu1DiffSCost}\end{subfigure}
	\begin{subfigure}{0.45\textwidth}\centering\includegraphics[scale=0.4]{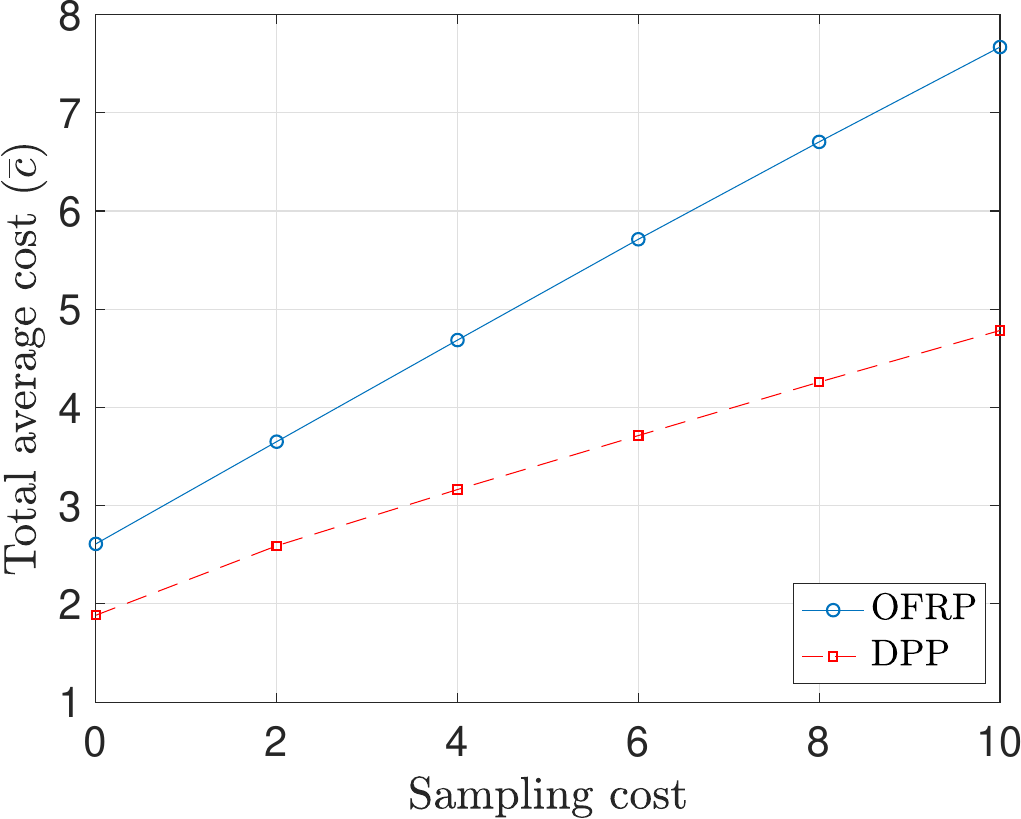}\caption{Performance comparison.}\label{Fig:PerfDiffSamplCost}\end{subfigure}
	\caption{Scheduling decisions and performance comparison for different values of the sampling cost, $c_\text{tr}=5$, $A=5$, $M=10$.}
	\label{Fig: ScheDecisionsSampl}
\end{figure}
\begin{figure}[t!]
	\centering
	\begin{subfigure}{0.45\textwidth}\centering\includegraphics[scale=0.44]{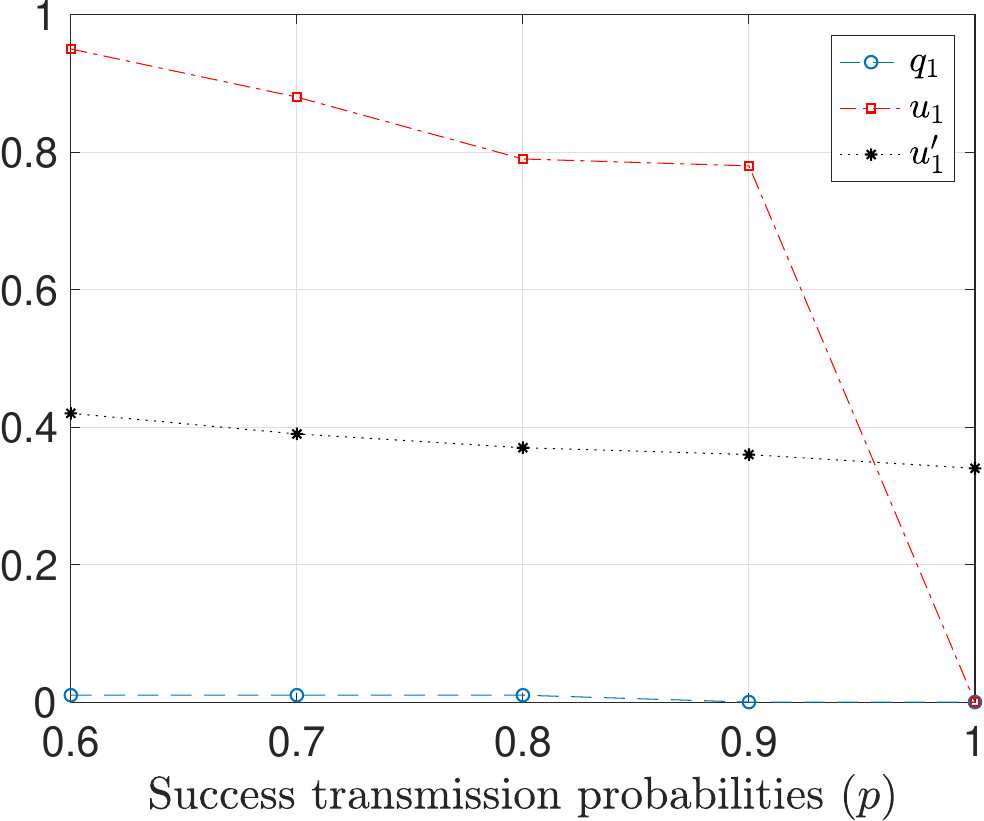}\caption{Optimized probabilities of the \ac{OFRP} policy.}\label{Fig:OptProbabilitiesDiffProb}\end{subfigure}
	\begin{subfigure}{0.46\textwidth}\centering\includegraphics[scale=0.44]{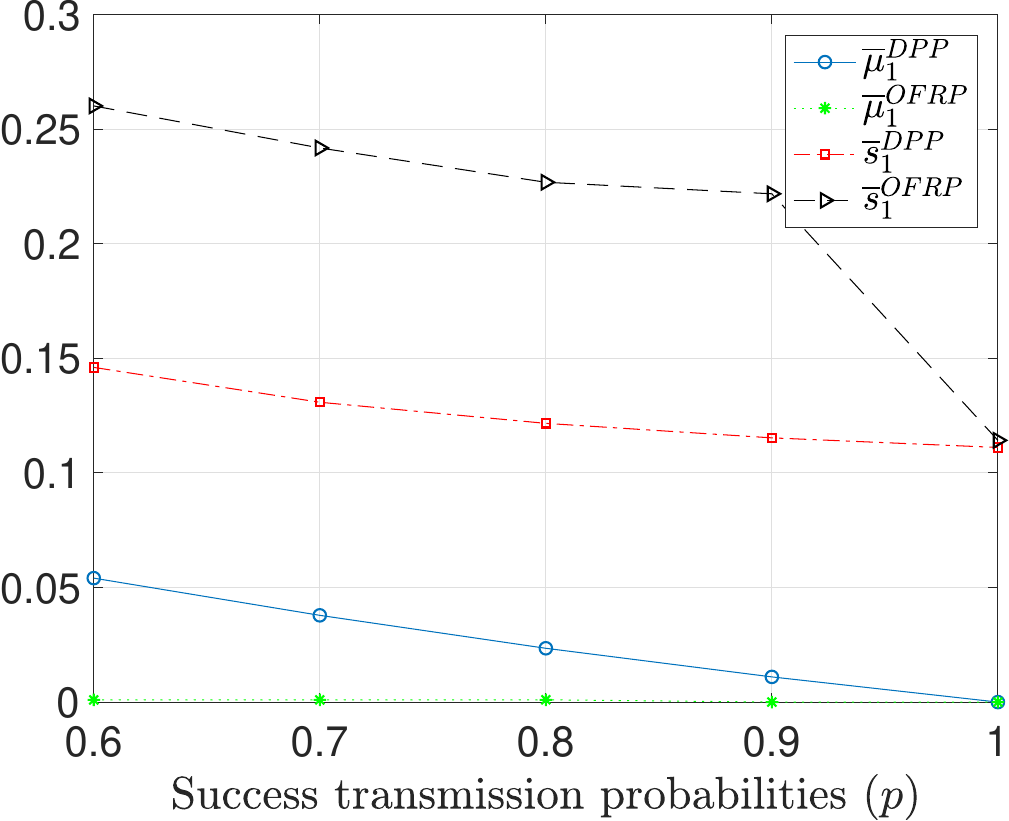}\caption{Average values of $\mu_1$ and $s_1$.}\label{Fig:AvValuesS1mu1DiffProb}\end{subfigure}
	\begin{subfigure}{0.45\textwidth}\centering\includegraphics[scale=0.44]{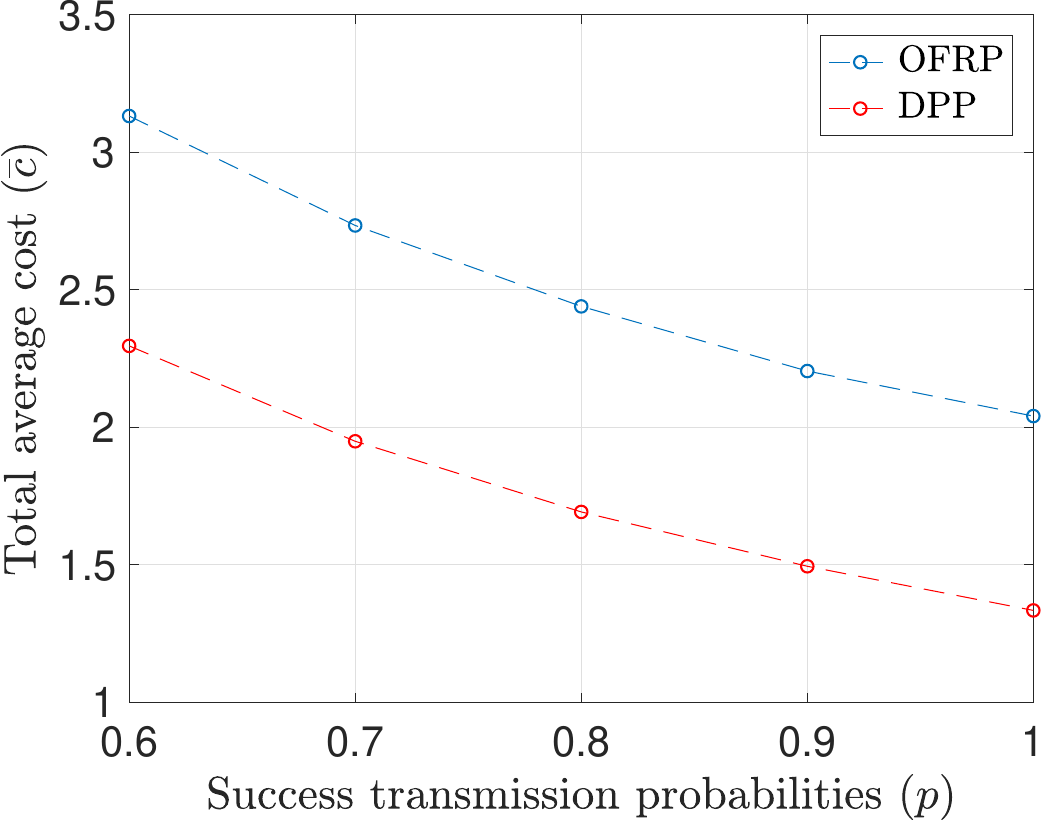}\caption{Total average cost.}\label{Fig:PerfDiffProb}\end{subfigure}
	\caption{Scheduling decisions for different values of the success probabilities. $A =5$, $M=10$, $c_\text{tr}=5$.}
	\label{Fig: ScheDecisionsProb}
\end{figure}
In Figures \ref{Fig: ScheDecisionsSampl} and \ref{Fig: ScheDecisionsProb}, we provide results that show the scheduling decisions of the scheduling policies for different values of the sampling and transmission cost. We consider a system with two symmetric users. More specifically, we consider that the users have the same success transmission probabilities, the same value of AoI constraints and the same transmission and sampling costs. We observe that in this case, the optimized scheduling probabilities are $\alpha_1=\alpha_2 = 0.5$. In these plots, we focus on the first user (user 1) to show the behavior of the policies for different transmission policies. In the figures with total cost, we depict the total average cost for both users. In Fig. \ref{Fig: ScheDecisionsSampl}, we consider that the transmission cost is $c_\text{tr} =5$, and the \ac{AoI} constraints  $A = 5$, where  $M=10$. Fig. \ref{Fig:OptProbabilitiesDiffSCost} depicts the optimized probabilities of the \ac{OFRP} policy for different values of the sampling cost. We observe that as the sampling cost takes values larger than the transmission cost, $u_1$  decreases  and $q_1$  increases. Note that these probabilities are conditional probabilities. In particular, $q_1$ is the probability to schedule user $1$ to transmit an old packet given that there is an old packet in the cache, and $u_1$ is the probability to  schedule user $1$ to transmit a fresh packet given that there is an old packet in the cache. 

In Fig. \ref{Fig:AvValuesS1mu1DiffSCost},  we compare the values of $\overline{\mu}_1$ and $\overline{s}_1$ of the \ac{DPP} policy with those of the \ac{OFRP} policy. We observe that $\overline{s}_1^{\text{DPP}} <\overline{s}_1^{\text{OFRP}}$ and $\overline{\mu}_1^{\text{DPP}}>\overline{\mu}_1^{\text{OFRP}}$. Therefore, with \ac{DPP} the user transmits old packets more frequently than \ac{OFRP}. Thus, the \ac{DPP} utilizes more efficiently the option of transmitting an old packet than \ac{OFRP}, and as a result, the corresponding cost is smaller for the case of the \ac{DPP} policy, as shown in Fig. \ref{Fig:PerfDiffSamplCost}.

In Fig. \ref{Fig: ScheDecisionsProb}, we provide results that show the behavior of the \ac{DPP} and \ac{OFRP} policies regarding the scheduling decisions for different values of the success probabilities. The values of the \ac{AoI} are $A  = 5$, and the transmission and sampling costs are $c_\text{tr}=5$ and $c_\text{s}=1$, respectively. Fig. \ref{Fig: ScheDecisionsProb} depicts the scheduling probabilities for the \ac{OFRP} policy. Although the sampling cost is significantly smaller than the transmission cost, we observe that for small values of $p$, the \ac{OFRP} utilizes the option of transmitting an old packet. The reason is that the sampling cost is small but not negligible. We also see that the \ac{DPP} policy utilizes more efficiently the option of transmitting an old packet than the \ac{OFRP}, as shown in Fig. \ref{Fig:AvValuesS1mu1DiffProb}. As a result, the performance of the \ac{DPP} policy  is better than that of the \ac{OFRP} policy, as shown in Fig. \ref{Fig:PerfDiffProb}.

\section{Conclusions}
In this work, we consider the total cost minimization problem while guaranteeing the required freshness of the data at the receiver. We propose three scheduling policies; one dynamic policy that optimizes the decisions slot by slot, and two stationary randomized policies of which the probabilities remain stable during the time. Simulation results show the importance of having the option to transmit an old packet instead of sampling a fresh each time that a user is scheduled. In particular, for the cases in which the sampling cost is larger than the transmission cost, we observe that by transmitting old packet we can significantly improve the total average cost while satisfying the average \ac{AoI} constraints.

\begin{appendices}
\section{Upper Bound on The Drift}
By	using the fact that $\left( \max[Q-b,0]+A \right)^2 \leq Q^2 +A^2 +b^2+2Q(A-b)$, we can rewrite \eqref{vqevolution} as 
$
	X^2_{k}(t+1)  \leq X_{k}^2(t) + A_{k}^2(t+1) +(A_{k}^{\text{max}})^2  
	+ 2X_{k}(t) (A_{k}(t+1)-(A_{k}^{\text{max}}))\text{.}
$
By rearranging the terms  and dividing by $2$, we obtain
\begin{align}\label{eq2A}
	\sum\limits_{k=1}^{K} X_{k}^2(t+1) - X^2_{k}(t) 
	\leq \sum\limits_{k=1}^K \frac{A_{k}^{2}(t+1) + (A_{k}^{\text{max}})^2 + 2X_{k}(t)(A_{k}(t+1)-A_{k}^{\text{max}})}{2}\text{.}
\end{align}
By taking the conditional expectation in \eqref{eq2A}, we obtain
\begin{align}\label{eq3A}
	\Delta(\mathbf{X}(t)) \leq \sum\limits_{k=1}^K \frac{\mathbb{E}\{A_k^2(t+1) | S_t\} + (A_k^{\text{max}})^2}{2}
	+ \sum\limits_{k=1}^K X_k(t)  \left[ \mathbb{E}\{A_k(t+1) | S_t\} -A_k^{\text{max}}\right]\text{.}
\end{align}
In order to proceed, we have to calculate $\mathbb{E} \{ A_k(t+1) | S_t\}$. By utilizing the expression for the evolution of \ac{AoI} in \eqref{eq: EvolutionAoI2}, we obtain
\begin{align}\nonumber
	&\mathbb{E} \{A_k(t+1) | S_t\} = \mathbb{E} \{(A_k^p(t)+1)d_k(t+1)
	+ (1-d_k(t+1)) \min\{A_k(t)+1,M \}  | S_t \}\\\label{eq: ExpAoI}
	& = \mathbb{E} \{ p_k\mu_k(t) + p_k s_k(t)|S_t \} (A^p_k(t) +1 ) 
	+ \mathbb{E} \{1-p_k\mu_k(t)  - p_ks_k(t) |S_t \}\min\{A_k(t)+1,M\}\text{.}
\end{align}
By substituting  \eqref{eq: ExpAoI} into \eqref{eq3A} and adding the term  $V\mathbb{E}\{c(t)|S_t\}$ on both sides, we obtain the result below
\begin{align}\nonumber
		&\Delta(\mathbf{X}(t)) + V\mathbb{E}\{c(t)|S_t\} \leq \sum\limits_{k=1}^K\frac{\mathbb{E}\{(A^2_{k}(t+1))|S_t\}+(A_{k}^{\text{max}})^2}{2}
		+\sum\limits_{k=1}^N\mathbb{E}\{X_{k}(t)[(A_{k}^p(t)+1)(p_k s_{k}(t)+ p_k\mu_{k}(t))\\
		&+\min\{(A_{k}(t)+1), M\}(1-p_{k}s_{k} (t) 
		- p_k\mu_{k}(t))] |S_t\} - X_{k}(t)A_{k}^{\text{max}}+ V\mathbb{E}\{c(t)|S_t\} \text{.}\label{eq28}
\end{align}
By setting $W_k(t) = p_k s_k(t)  + p_k\mu_k(t)$, we obtain the result in \eqref{eq: DriftBound}.
\rev{
\section{Proof of Theorem 2}
\begin{proof}
Since the \ac{DPP} algorithm uses the method of oppurtunistically minimizing an expectation, we get
\begin{align}\nonumber
	&\Delta(\mathbf{X}(t)) + V\mathbb{E}\{c(t)|S_t\}  
	\leq B + \sum\limits_{k=1}^{K} X_k(t) \left[\right.\mathbb{E}\{W_k(t)(A_k^p(t)+1)  + (1-W_k(t))\times\\
	& \min\left[A_{k}(t)+1, M\right]| S_t\}
	-A_k^{\text{max}} \left.\right]   + V\mathbb{E}\{c(t)|S_t\} \leq B + \sum\limits_{k=1}^K X_k(t) \mathbb{E}\{y^*_k(t)\} + V\mathbb{E}\{c^*(t)\}\text{,}
\end{align}
where $y^*(t)$ and $c^*(t)$ are the resulting vaues of $\omega$ policy as defined in Lemma 2. Note that the $\omega$ policy is independent of the system state $S_t$.
By utilizing the bounds of Lemma 2, i.e., bound $\eqref{eq:costbound}$, we get
$
 \Delta(\mathbf{X}(t)) + \mathbb{E} \{c(t)|S(t)\}\leq B + \epsilon \sum\limits_{k=1}^K X_k(t) + V(c^{\text{opt}} + \epsilon)\text{,}
$
by taking $\epsilon\rightarrow 0$, we have 
$	\Delta(\mathbf{X}(t)) + \mathbb{E} \{c(t)|S(t)\} \leq B + Vc^{\text{opt}}\text{,}
$
where is the exact form of the Lyapunov optimization theorem \cite[Theorem 4.2]{neely2010stochastic}. Therefore, all virtual queues are mean rate stable, and therefore, the average constraints are satisfied. 
\end{proof}
\section{Proof of Theorem 3}
\begin{proof}
We apply the law of iterated expectations in equation \eqref{eq:driftdelta}, we obtain
\begin{align}
	\mathbb{E}\{L(\mathbf{X}(t+1))\} - \mathbb{E}\{L(\mathbf{X}(t))\} + V\mathbb{E}\{c(t)\} \leq 
	B + \epsilon \sum\limits_{k=1}^K X_k(t) + Vc^{\text{opt}}  + \epsilon\text{.}
\end{align}
Summing over $\tau = 0,1,2,\ldots t-1$, and diving by $t$ and $V$, we get
\begin{align}\nonumber
	\sum\limits_{\tau=0}^{t-1} \frac{1}{t} \mathbb{E}\{c(t)\} \leq \frac{\mathbb{E}\{L(\mathbf{X}(t))\} - \mathbb{E}\{L(\mathbf{X}(0))\} }{Vt} + \frac{B}{V} + \frac{\epsilon}{V}\frac{1}{t} \sum\limits_{k=1}^K X_k(t) + c^{\text{opt}} + \frac{\epsilon}{V}\text{.}
\end{align}
By taking $\epsilon \rightarrow 0$, and $t \rightarrow \infty$, we get, 
$
	\lim\limits_{t\rightarrow \infty}\sup \frac{1}{t} \sum\limits_{\tau=0}^{t-1} \mathbb{E}\{c(t)\} \leq \frac{B}{V} + c^{\text{opt}}\text{.}
$
This concludes the result.
\end{proof}}
\end{appendices}
 
\bibliography{MyBibAge}

\begin{thebibliography}{10}

\bibitem{FountoulakisOptimalSampling}
E.~{Fountoulakis}, N.~{Pappas}, M.~{Codreanu}, and A.~{Ephremides}, ``Optimal
  sampling cost in wireless networks with age of information constraints,''
  {\em in Proc. IEEE INFOCOM}, pp.~918--923, June 2020.

\bibitem{kostamonograph2017age}
A.~Kosta, N.~Pappas, and V.~Angelakis, ``Age of information: A new concept,
  metric, and tool,'' {\em Foundations and Trends{\textregistered} in
  Networking}, vol.~12, no.~3, pp.~162--259, 2017.

\bibitem{sunmodiano2019age}
Y.~Sun, I.~Kadota, R.~Talak, and E.~Modiano, ``Age of information: A new metric
  for information freshness,'' {\em Synthesis Lectures on Communication
  Networks}, vol.~12, no.~2, pp.~1--224, 2019.

\bibitem{yates2020age}
R.~D. {Yates}, Y.~{Sun}, D.~{Richard Brown}, S.~K. {Kaul}, E.~{Modiano}, and
  S.~{Ulukus}, ``Age of information: An introduction and survey,'' {\em IEEE
  Journal on Selected Areas in Communications}, pp.~1--1, 2021.

\bibitem{kaulyates2012real}
S.~Kaul, R.~Yates, and M.~Gruteser, ``Real-time status: How often should one
  update?,'' {\em in Proc. IEEE INFOCOM}, pp.~2731--2735, Mar. 2012.

\bibitem{kamephr2015effect}
C.~Kam, S.~Kompella, G.~D. Nguyen, and A.~Ephremides, ``Effect of message
  transmission path diversity on status age,'' {\em IEEE Trans. Inf. Theory},
  vol.~62, no.~3, pp.~1360--1374, Mar. 2016.

\bibitem{kosta2019queuemanage}
A.~Kosta, N.~Pappas, A.~Ephremides, and V.~Angelakis, ``Age of information
  performance of multiaccess strategies with packet management,'' {\em Journal
  of Commun. and Networks}, vol.~21, no.~3, pp.~244--255, June 2019.

\bibitem{talak2020age}
R.~Talak and E.~Modiano, ``Age-delay tradeoffs in queueing systems,'' {\em IEEE
  Trans. Inf. Theory}, 2020.

\bibitem{pappas2019delay}
N.~Pappas and M.~Kountouris, ``Delay violation probability and age of
  information interplay in the two-user multiple access channel,'' {\em in
  Proc. IEEE SPAWC}, pp.~1--5, 2019.

\bibitem{fountoulakis2020information}
E.~Fountoulakis, T.~Charalambous, N.~Nomikos, and N.~Pappas, ``Information
  freshness and packet drop rate interplay in a two-user multi-access
  channel,'' {\em accepted on ITW}, 2021.

\bibitem{maatouk2020age}
A.~Maatouk, M.~Assaad, and A.~Ephremides, ``On the age of information in a csma
  environment,'' {\em IEEE/ACM Trans. Net}, vol.~28, no.~2, pp.~818--831, 2020.

\bibitem{kadota2021minimizing}
I.~Kadota and E.~Modiano, ``Minimizing the age of information in wireless
  networks with stochastic arrivals,'' {\em IEEE Trans. Mobil. Comput.},
  vol.~20, no.~3, pp.~1173--1185, 2021.

\bibitem{joo2018wireless}
C.~Joo and A.~Eryilmaz, ``Wireless scheduling for information freshness and
  synchrony: Drift-based design and heavy-traffic analysis,'' {\em IEEE/ACM
  Trans. Net.}, vol.~26, no.~6, pp.~2556--2568, 2018.

\bibitem{qian2020minimizing}
Z.~Qian, F.~Wu, J.~Pan, K.~Srinivasan, and N.~B. Shroff, ``Minimizing age of
  information in multi-channel time-sensitive information update systems,''
  {\em in Proc. IEEE INFOCOM}, pp.~446--455, 2020.

\bibitem{talak2020improving}
R.~Talak, S.~Karaman, and E.~Modiano, ``Improving age of information in
  wireless networks with perfect channel state information,'' {\em IEEE/ACM
  Trans. Net.}, vol.~28, no.~4, pp.~1765--1778, 2020.

\bibitem{chen2019optimal}
Z.~{Chen}, N.~{Pappas}, E.~{Björnson}, and E.~G. {Larsson}, ``Optimizing
  information freshness in a multiple access channel with heterogeneous
  devices,'' {\em IEEE Open Journal of the Communications Society}, vol.~2,
  pp.~456--470, 2021.

\bibitem{moltafet2019power}
M.~Moltafet, M.~Leinonen, M.~Codreanu, and N.~Pappas, ``Power minimization for
  age of information constrained dynamic control in wireless sensor networks,''
  {\em IEEE Trans. Commun.}, vol.~69, no.~2, pp.~1121--1133, 2021.

\bibitem{tripathi2019whittle}
V.~Tripathi and E.~Modiano, ``A whittle index approach to minimizing functions
  of age of information,'' {\em in Proc. ASILOMAR}, pp.~1160--1167, Sept. 2019.

\bibitem{ceran2019reinforcement}
E.~T. Ceran, D.~G{\"u}nd{\"u}z, and A.~Gy{\"o}rgy, ``Reinforcement learning to
  minimize age of information with an energy harvesting sensor with {HARQ} and
  sensing cost,'' {\em in Proc. IEEE INFOCOM}, pp.~656--661, Apr. 2019.

\bibitem{abd2020aoi}
M.~A. Abd-Elmagid, H.~S. Dhillon, and N.~Pappas, ``Ao{I}-optimal joint sampling
  and updating for wireless powered communication systems,'' {\em IEEE Trans
  Vehic. Tech.}, vol.~69, no.~11, pp.~14110--14115, 2020.

\bibitem{zhou2019joint}
B.~Zhou and W.~Saad, ``Joint status sampling and updating for minimizing age of
  information in the internet of things,'' {\em IEEE Trans. Commun.}, vol.~67,
  no.~11, pp.~7468--7482, 2019.

\bibitem{StamatakisIoT}
G.~{Stamatakis}, N.~{Pappas}, and A.~{Traganitis}, ``Optimal policies for
  status update generation in an iot device with heterogeneous traffic,'' {\em
  IEEE Internet of Things Journal}, vol.~7, no.~6, pp.~5315--5328, 2020.

\bibitem{kadota2019scheduling}
I.~Kadota, A.~Sinha, and E.~Modiano, ``Scheduling algorithms for optimizing age
  of information in wireless networks with throughput constraints,'' {\em
  IEEE/ACM Trans. Net.}, vol.~27, no.~4, pp.~1359--1372, 2019.

\bibitem{BedewyOptimalSampling}
A.~M. {Bedewy}, Y.~{Sun}, S.~{Kompella}, and N.~B. {Shroff}, ``Optimal sampling
  and scheduling for timely status updates in multi-source networks,'' {\em
  IEEE Trans. Inf. Theory}, pp.~1--1, 2021.

\bibitem{yao2020age}
G.~Yao, A.~M. Bedewy, and N.~B. Shroff, ``Age minimization transmission
  scheduling over time-correlated fading channel under an average energy
  constraint,'' {\em arXiv preprint arXiv:2012.02958}, 2020.

\bibitem{fountoulakis2021dynamic}
E.~Fountoulakis, T.~Charalambous, A.~Ephremides, and N.~Pappas, ``A dynamic
  scheduling policy for a network with heterogeneous time-sensitive traffic,''
  {\em arXiv preprint arXiv:2109.04784}, 2021.

\bibitem{huang2020real}
K.~Huang, W.~Liu, M.~Shirvanimoghaddam, Y.~Li, and B.~Vucetic, ``Real-time
  remote estimation with hybrid {ARQ} in wireless networked control,'' {\em
  IEEE Trans. Wireless Commun.}, vol.~19, no.~5, pp.~3490--3504, 2020.

\bibitem{kam2020age}
C.~Kam, S.~Kompella, and A.~Ephremides, ``Age of incorrect information for
  remote estimation of a binary markov source,'' {\em in Proc. IEEE INFOCOM
  Workshops}, pp.~1--6, Jul. 2020.

\bibitem{inoue2019aoi}
Y.~Inoue and T.~Takine, ``Ao{I} perspective on the accuracy of monitoring
  systems for continuous-time markovian sources,'' {\em in Proc. IEEE INFOCOM
  Workshops}, pp.~183--188, Apr. 2019.

\bibitem{ornee2019sampling}
T.~Z. Ornee and Y.~Sun, ``Sampling for remote estimation through queues: Age of
  information and beyond,'' {\em in Proc. WiOpt}, pp.~1--8, Jun. 2019.

\bibitem{huang2019retransmit}
K.~Huang, W.~Liu, Y.~Li, and B.~Vucetic, ``To retransmit or not: Real-time
  remote estimation in wireless networked control,'' {\em in Proc. IEEE ICC},
  pp.~1--7, May 2019.

\bibitem{kountouris2020semantics}
M.~Kountouris and N.~Pappas, ``Semantics-empowered communication for networked
  intelligent systems,'' {\em Accepted, to appear, IEEE Commun. Magazine},
  2021.

\bibitem{maatouk2020ageSemantics}
A.~Maatouk, M.~Assaad, and A.~Ephremides, ``The age of incorrect information:
  an enabler of semantics-empowered communication,'' {\em arXiv preprint
  arXiv:2012.13214}, 2020.

\bibitem{popovski2020semantic}
P.~Popovski, O.~Simeone, F.~Boccardi, D.~G{\"u}nd{\"u}z, and O.~Sahin,
  ``Semantic-effectiveness filtering and control for post-5{G} wireless
  connectivity,'' {\em Journal of the Indian Institute of Science}, vol.~100,
  no.~2, pp.~435--443, 2020.

\bibitem{ceran2021reinforcement}
E.~T. Ceran, D.~G{\"u}nd{\"u}z, and A.~Gy{\"o}rgy, ``A reinforcement learning
  approach to age of information in multi-user networks with harq,'' {\em IEEE
  Journal on Selected Areas in Communications}, vol.~39, no.~5, pp.~1412--1426,
  2021.

\bibitem{maatouk2020optimality}
A.~Maatouk, S.~Kriouile, M.~Assad, and A.~Ephremides, ``On the optimality of
  the whittle’s index policy for minimizing the age of information,'' {\em
  IEEE Transactions on Wireless Communications}, vol.~20, no.~2,
  pp.~1263--1277, 2020.

\bibitem{hatami2021aoi}
M.~Hatami, M.~Leinonen, and M.~Codreanu, ``Aoi minimization in status update
  control with energy harvesting sensors,'' {\em IEEE Trans. Comm.}, vol.~69,
  no.~12, pp.~8335--8351, 2021.

\bibitem{neely2010stochastic}
M.~J. Neely, ``Stochastic network optimization with application to
  communication and queueing systems,'' {\em Synthesis Lectures on
  Communication Networks}, vol.~3, no.~1, pp.~1--211, 2010.

\bibitem{neely2010queue}
M.~J. Neely, ``Queue stability and probability 1 convergence via {L}yapunov
  optimization,'' {\em arXiv preprint arXiv:1008.3519}, 2010.

\bibitem{MeynMarkovChains}
S.~Meyn and R.~L. Tweedie, {\em Markov Chains and Stochastic Stability}.
\newblock 2nd ed. New York, NY, USA: Cambridge University Press, 2010.

\end{thebibliography}
\bibliographystyle{ieeetr}

\end{document}